\documentclass[preprint,aps,pra,showpacs,floatfix]{revtex4-1}
\usepackage{graphicx}
\usepackage{times}
\usepackage{nicefrac}
\usepackage{amsmath}
\usepackage{amsfonts}
\usepackage{amssymb}
\usepackage{amsthm}
\usepackage{epsf}
\usepackage{bm}
\usepackage{hyperref}
\usepackage{dcolumn}
\usepackage{tabularx}
\newcolumntype{.}{D{x}{}{-1}}
\usepackage{mathtools}
\usepackage{commath}
\usepackage{fancyhdr}
\usepackage{longtable}
\newcommand{\veps}{\varepsilon}

\newcommand{\ket}[1]{|#1\rangle}
\newcommand{\braket}[2]{\langle#1|#2\rangle}
\newcommand{\matrixel}[3]{\left\langle#1\left|\vphantom{#1}#2\vphantom{#3}\right|#3\right\rangle}
\newcommand{\matrixelb}[3]{\big\langle#1\big|#2\big|#3\big\rangle}
\newcommand{\balpha}{\bm{\alpha}}

\newcommand{\bfB}{{\bf B}}

\newcommand{\bfp}{{\bf p}}

\newcommand{\bfr}{{\bf r}}
\newcommand{\hbfr}{\hat{\bf r}}

\newcommand{\aZ}{\alpha Z}
\newcommand{\muB}{\mu_\textrm{B}}
\newcommand{\Vnuc}{V_\mathrm{nuc}}
\newcommand{\Vscr}{V_\mathrm{scr}}
\newcommand{\Vmagn}{V_\mathrm{m}}
\newcommand{\ta}{\tilde a}
\newcommand{\tb}{\tilde b}
\newcommand{\tveps}{\tilde \veps}
\newcommand{\tomega}{\tilde \omega}
\newcommand{\gBB}[1]{g^{(2)}_{#1}}
\newcommand{\gBBB}[1]{g^{(3)}_{#1}}
\newcommand{\gk}[2]{g^{(#1)}_{#2}}
\newcommand{\dgk}[2]{\Delta g^{(#1)}_{#2}}
\newcommand{\stgr}{{}^2P_{1/2}}
\newcommand{\stex}{{}^2P_{3/2}}
\newcommand{\stJ}{{}^2P_{J}}
\newcommand{\lstgr}{[(1s)^2\,(2s)^2\,2p]\,{}^2P_{1/2}}
\newcommand{\lstex}{[(1s)^2\,(2s)^2\,2p]\,{}^2P_{3/2}}
\newcommand{\gBBint}{\gBB{\mathrm{int}}}
\newcommand{\dgBBinto}{\Delta \gBB{\mathrm{int,1}}}
\newcommand{\gBBBint}{\gBBB{\mathrm{int}}}
\newcommand{\dgBBBinto}{\Delta \gBBB{\mathrm{int,1}}}
\newcommand{\dgkinto}[1]{\Delta \gk{#1}{\mathrm{int,1}}}
\newcommand{\dexp}[1]{\times 10^{#1}}
\begin{document}
\title{Interelectronic-interaction contribution to the nonlinear Zeeman effect in boronlike ions}
\author{
A.~S.~Varentsova,$^{1,2}$
V.~A.~Agababaev,$^{2,3,4}$
D.~A.~Glazov,$^{2,3}$
A.~M.~Volchkova,$^{2,3}$
A.~V.~Volotka,$^{3,5}$
V.~M.~Shabaev,$^{3}$
and G.~Plunien$^{6}$
}
\affiliation{
$^1$ ITMO University, Kronverksky pr. 49, 197101 St. Petersburg, Russia \\
$^2$ State Scientific Centre ``Institute for Theoretical and Experimental Physics'' of National Research Centre ``Kurchatov Institute'', B.~Cheremushkinskaya st. 25, 117218 Moscow, Russia \\
$^3$ Department of Physics, St. Petersburg State University, Universitetskaya 7/9, 199034 St. Petersburg, Russia \\
$^4$ St. Petersburg Electrotechnical University ``LETI'', Professor Popov st. 5, 197376 St. Petersburg, Russia \\
$^5$ Helmholtz-Institut Jena, Fr\"obelstieg 3, D-07743 Jena, Germany \\
$^6$ Institut f\"ur Theoretische Physik, Technische Universit\"at Dresden, Mommsenstra{\ss}e 13, D-01062 Dresden, Germany
}
\begin{abstract}
Relativistic calculations of the second- and third-order contributions in magnetic field to the Zeeman splitting in boronlike ions are presented for the wide range of nuclear charge numbers $Z=6$--$92$. The interelectronic-interaction correction of the first order in $1/Z$ is evaluated to all orders in $\aZ$. The higher-order corrections in $1/Z$ are taken into account approximately by means of effective screening potentials. The obtained results are important for interpretation of experimental data on the Zeeman splitting in boronlike ions, in particular, for the ARTEMIS experiment presently implemented at GSI.
\end{abstract}
\maketitle
%
\section{Introduction}
%
Investigations of the $g$ factor of highly charged ions can serve for stringent tests of the bound-state QED, the determination of fundamental constants and of nuclear parameters, see, e.g., Refs. \cite{sturm:13:ap,volotka:13:ap,volotka:14:prl,shabaev:15:jpcrd}. Meanwhile, the precision of the $g$-factor measurements in Penning traps has reached the level of $10^{-9}$--$10^{-11}$ \cite{sturm:13:ap,haeffner:00:prl,verdu:04:prl,sturm:11:prl,sturm:13:pra,wagner:13:prl,vogel:13:ap}. Achievement of the comparable theoretical accuracy requires consideration of the QED diagrams up to the second order in $\alpha$, higher-order correlation contributions, and various nuclear effects, see Ref.~\cite{shabaev:15:jpcrd} and more recent works~\cite{czarnecki:16:pra,yerokhin:17:pra-1,yerokhin:17:pra-2,shabaev:17:prl,malyshev:17:jetpl,czarnecki:18:prl,glazov:18:os}. The long-term combined experimental and theoretical efforts have provided the most accurate up-to-date value of the electron mass~\cite{sturm:14:n,CODATA14,zatorski:17:pra}. Recent measurement for two lithium-like calcium isotopes \cite{koehler:16:nc} is sensitive to the relativistic nuclear recoil effect indicating the potential to access bound-state QED effects beyond the Furry picture \cite{shabaev:17:prl,malyshev:17:jetpl}. Further investigations with few-electron ions can provide an independent determination of the fine structure constant \cite{shabaev:06:prl,volotka:14:prl-np,yerokhin:16:prl}. 

The nonlinear effects in magnetic field play an important role in boronlike ions \cite{lindenfels:13:pra,glazov:13:ps}. Due to the mirror symmetry, the second-order effect does not influence the ground-state Zeeman splitting in hydrogenlike, lithiumlike, and boronlike ions. However, it becomes observable in the Zeeman splitting of the $\stex$ state of boronlike ions, which can be measured by the laser-microwave double-resonance spectroscopy~\cite{quint:08:pra,lindenfels:13:pra}. Moreover, the quadratic effect gives a tiny correction to the fine-structure transition energies, which are measured with an increasing precision \cite{draganic:03:prl,soriaorts:06:prl,maeckel:11:prl,maeckel:13:ps}. The third-order effect contributes to the Zeeman splitting of any state, although in hydrogenlike and lithiumlike ions it is negligible at the present level of accuracy. In boronlike ions the nonlinear contributions are strongly enhanced due to the mixing of the fine-structure components. 

The first high-precision $g$-factor measurement in boronlike ion was performed for $\mathrm{Ar}^{13+}$ at the MPIK with the laser spectroscopy \cite{soriaorts:07:pra}. The ARTEMIS experiment at GSI implements the laser-microwave double-resonance technique to reach the ppb precision for the $g$ factors of the ground $\stgr$ and the first excited $\stex$ states in boronlike argon \cite{lindenfels:13:pra,vogel:15:hi}. The new Penning-trap experiment at the MPIK in Heidelberg, called ALPHATRAP \cite{alphatrap}, primarily aims at the $g$ factor of heavy few-electron ions, including boronlike. 

Since the discovery of the quadratic Zeeman effect \cite{jenkins:39:pr,schiff:39:pr} numerous experimental studies in atoms and molecules have been accomplished \cite{garton:69:aj,raoult:05:jpb,pohl:09:prep,numazaki:10:pra}. The nonlinear Zeeman effect is important in astrophysics due to the strong magnetic field at the stellar objects \cite{preston:70:aj,moran:98:ras,zakharov:04:na}. It is also relevant for atomic clocks, magnetometers and other high-precision experimental setups as a source of the systematic shifts \cite{numazaki:10:pra,ludlow:15:rmp,hu:17:pra,bao:18:prl}. 

The relativistic theory of the second-order effect was developed, e.g., in Refs.~\cite{manakov:74:jpb,manakov:76:pla,grozdanov:86:jpb,feinberg:90:pra,szmytkowski:02:jpb,szmytkowski:02:pra}. However, much less results are available for the third-order contribution, apart from the all-order calculations \cite{chen:92:pra,rutkowski:05:ps,nakashima:10:aj,rozenbaum:14:pra}. Most of these  investigations were devoted to hydrogen and hydrogen-like ions. In our recent study~\cite{agababaev:17:nimb} the leading-order interelectronic-interaction and QED corrections to second-order Zeeman splitting have been evaluated for boronlike argon. Relativistic calculations of the cubic effect have been performed in Ref.~\cite{varentsova:17:nimb} for hydrogenlike, lithiumlike, and boronlike ions. The results obtained within the independent-electron approximation utilizing different screening potentials have shown the need for the rigorous treatment of the interelectronic interaction. In the present paper, we evaluate the one-photon-exchange correction to the second- and third-order Zeeman effects for the ground $\stgr$ and the first excited $\stex$ states of boronlike ions for the range of nuclear charge numbers $Z=6$--$92$. An efficient numerical method is developed to treat the contributions of arbitrary order in the magnetic field. The calculations are exact to all orders in $\aZ$ in the zeroth and first orders in $1/Z$, and partially include higher orders in $1/Z$ by the use of the screening potentials.

The relativistic units ($\hbar = 1$, $m_e = 1$, $c = 1$) and the Heaviside charge unit ($\alpha=e^2/(4\pi), e<0$) are used throughout the paper.
%
\section{Description of the method}
%
\subsection{Zeeman effect}
%
\begin{figure}
  \includegraphics[width=\textwidth]{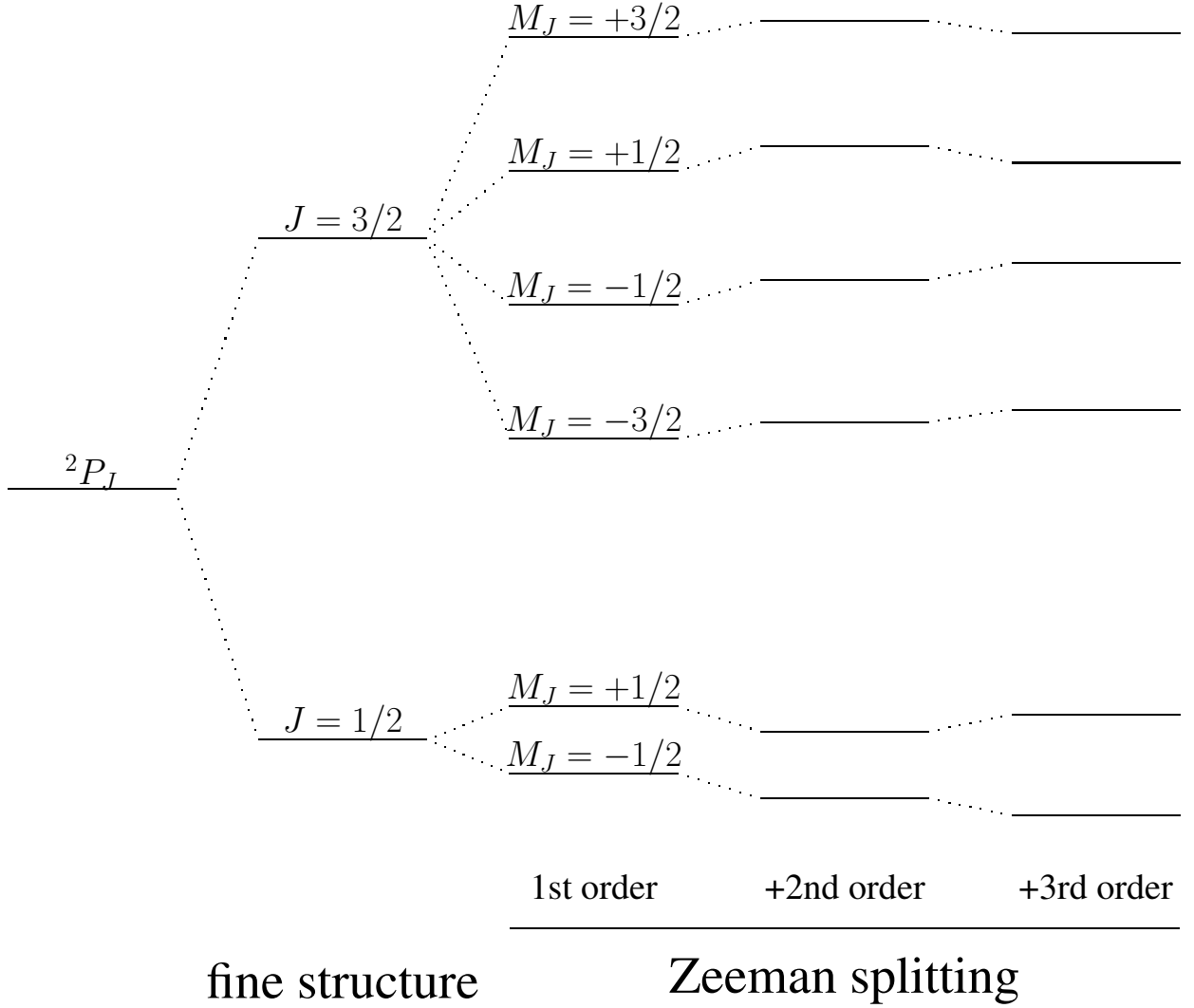}
  \caption{Level structure of the $\stJ$ states of boronlike ion in the presence of magnetic field (not to scale).
  \label{fig:levels}}
\end{figure}
We consider a five-electron ion in the ground $\lstgr$ or in the first excited $\lstex$ state. In the presence of an external magnetic field $\bfB$ the energy of an individual level $\ket{A}=\ket{J,M_J}$ with the total angular momentum $J$ and projection $M_J$ can be written as
\begin{equation}
  E_A (B) = E_A^{(0)} + \Delta E_A^{(1)} (B) + \Delta E_A^{(2)} (B) + \Delta E_A^{(3)} (B) + \dots
\,,
\end{equation}
where each term $\Delta E_A^{(k)} (B)$ is proportional to $B^k$. The first-order Zeeman effect is expressed through the $g$ factor,
\begin{equation}
  \Delta E_A^{(1)} (B) = g M_J \muB B
\,,
\end{equation}
where $\muB$ is the Bohr magneton. The analogous dimensionless coefficients for the second- and third-order Zeeman effect can be defined as \cite{lindenfels:13:pra}
\begin{align}
\label{eq:g-2}
  \Delta E_A^{(2)} (B) &= \gBB{}(M_J) (\muB B)^2
\,,\\
\label{eq:g-3}
  \Delta E_A^{(3)} (B) &= \gBBB{}(M_J) (\muB B)^3
\,.
\end{align}
Due to the symmetry relation $\gBB{}(-M_j) = \gBB{}(M_j)$, the second-order term does not contribute to the ground-state Zeeman splitting in hydrogenlike, lithiumlike, and boronlike ions. However, it alters the Zeeman splitting of the $\stex$ state and the fine-structure interval. For the third-order term we have $\gBBB{}(-M_j) = -\gBBB{}(M_j)$, so it contributes to the Zeeman splitting for any state and can be considered as a field-dependent correction to the $g$ factor. The scheme of the individual levels for both $\stJ$ states presented in Fig.~\ref{fig:levels} demonstrates the influence of the first-, second-, and third-order effects.

Within the independent-electron approximation, the Zeeman splitting for the ground $\stgr$ and the first excited $\stex$ state of the boronlike ion is determined by the valence electron only. Let $\ket{a}$ and $\veps_a$ be the wave function and energy of the valence electron in the $2p_{j}$ state, $\ket{b}$ and $\veps_b$ be the $1s$ or $2s$ core-electron wave function and energy, in the absence of external magnetic field. These unperturbed single-particle wave functions and energies are determined by the Dirac equation,
\begin{equation}
\label{eq:D}
\left[ \balpha \cdot \bfp + \beta + \Vnuc(r) + \Vscr(r) \right] \ket{a} = \veps_a \ket{a} 
\,,
\end{equation}
and similar for $\ket{b}$ and $\veps_b$. Here $\balpha$ and $\beta$ are the Dirac matrices, $\Vnuc(r)$ is the nuclear potential, $\Vscr(r)$ is the effective local screening potential employed to describe the interelectronic interaction within the independent-electron approximation.

The interaction with the external homogeneous magnetic field $\bfB$ is described by the operator 
\begin{equation}
\label{eq:Vmagn}
  \Vmagn = \muB \bfB \cdot [\bfr \times \balpha]
\,.
\end{equation}
Assuming that $\bfB$ is directed along the $z$ axis, we rewrite this as 
\begin{equation}
\label{eq:Vmagn-U}
  \Vmagn = \lambda U
\,,\qquad
  \lambda = \muB B
\,,\qquad
  U = [\bfr \times \balpha]_z
\,.
\end{equation}
It is convenient now to use $\lambda$ as the expansion parameter of the perturbation theory. 

The exact wave functions $\ket{\ta}$ and energies $\tveps_a$ in the presence of magnetic field satisfy the following Dirac equation,
\begin{equation}
\label{eq:D-m}
\left[ \balpha \cdot \bfp + \beta + \Vnuc(r) + \Vscr(r) + \Vmagn \right] \ket{\ta} = \tveps_a \ket{\ta}
\,,
\end{equation}
and similar for $\ket{\tb}$ and $\tveps_b$. Within the perturbation theory with respect to $\Vmagn$, they are given by the expansions
\begin{equation}
\label{eq:ten}
  \tveps_a = \sum_{k=0}^{\infty} \lambda^k \veps_a^{(k)}
\end{equation}
and
\begin{equation}
\label{eq:ta}
  \ket{\ta}
    = \sum_{k=0}^{\infty} \lambda^k \ket{a^{(k)}}
    = \sum_{k=0}^{\infty} \lambda^k \sum_n \ket{n} \braket{n}{a^{(k)}}
\,.
\end{equation}
Here $\ket{n}$ are the unperturbed wave functions from the spectrum of the Dirac equation~(\ref{eq:D}). Zeroth-order term is the unperturbed energy, $\veps_a^{(0)}=\veps_a$, and the first-order term is related to the $g$ factor: $\veps_a^{(1)}=g M_J$. While Eq.~(\ref{eq:D-m}) determines the so-called Dirac value of the $g$ factor, there are various corrections due to the interelectronic-interaction, radiative, and nuclear effects (see, e.g., Refs.~\cite{volotka:13:ap,shabaev:15:jpcrd} for a review). The second-order term $\veps_a^{(2)}$ yields the leading contribution to $\gBB{}$, which we denote as $\gBB{0}\equiv\veps_a^{(2)}$. There is also a nonzero contribution of the closed-shell electrons to $\gBB{}$, which is independent of the valence-electron state, so it influences neither the Zeeman splitting, nor the fine-structure splitting of the $\stJ$ states under consideration. For this reason, we neglect it in the present work as was done in Refs.~\cite{lindenfels:13:pra,glazov:13:ps,agababaev:17:nimb}. The third-order term, $\gBBB{0}\equiv\veps_a^{(3)}$, has no closed-shell counterpart.
%
\subsection{Perturbation theory}
%
In order to evaluate efficiently the higher-order contributions, we use the recursive representation of the Rayleigh-Schr\"odinger perturbation theory \cite{messiah}. Insertion of the expansions (\ref{eq:ten}) and (\ref{eq:ta}) into the Dirac equation (\ref{eq:D-m}) and separation of the terms of given order $k$ lead to the following recursive system of equations for the coefficients $\veps_a^{(k)}$ and $\braket{n}{a^{(k)}}$,
\begin{align}
\label{eq:rec-en}
  \veps_a^{(k)} &= \sum_m \matrixel{a}{U}{m} \braket{m}{a^{(k-1)}}
    - \sum_{j=1}^{k-1} \veps_a^{(j)} \braket{a}{a^{(k-j)}}
\,,\\
\label{eq:rec-na}
  {\left. \braket{n}{a^{(k)}} \right|}_{n \neq a} &= \frac{1}{\veps_a - \veps_n}
    \left[ \sum_m \matrixel{n}{U}{m} \braket{m}{a^{(k-1)}}
    - \sum_{j=1}^{k-1} \veps_a^{(j)} \braket{n}{a^{(k-j)}} \right]
\,,\\
\label{eq:rec-aa}
  \braket{a}{a^{(k)}} &= - \frac{1}{2}\,\sum_{j=1}^{k-1}
    \sum_m \braket{a^{(j)}}{m} \braket{m}{a^{(k-j)}}
\,,
\end{align}
with the initial values,
\begin{align}
  \veps_a^{(0)} = \veps_a
\,,\qquad
  {\left. \braket{n}{a^{(0)}} \right|}_{n \neq a} = 0
\,,\qquad
  \braket{a}{a^{(0)}} = 1
\,.
\end{align}
Equation~(\ref{eq:rec-aa}) is the consequence of the normalisation condition $\braket{\ta}{\ta}=1$. All the formulas presented for the energy $\tveps_a$ and the wave function $\ket{\ta}$ of the valence-electron state can be rewritten for the closed-shell states as well. One can easily check that this recursive system leads to the well-known expressions for the leading terms,
\begin{align}
\label{eq:e1}
  \veps_a^{(1)} =& \matrixel{a}{U}{a}
\,,\\
\label{eq:e2}
  \veps_a^{(2)} =& {\sum_n}' \frac{\matrixel{a}{U}{n}\matrixel{n}{U}{a}}{\veps_a - \veps_n}
\,,\\
\label{eq:e3}
  \veps_a^{(3)} =& {\sum_{n_1,n_2}}' \frac{\matrixel{a}{U}{n_1}\matrixel{n_1}{U}{n_2}\matrixel{n_2}{U}{a}}{(\veps_a - \veps_{n_1})(\veps_a - \veps_{n_2})}
    - {\sum_n}' \frac{\matrixel{a}{U}{n}\matrixel{n}{U}{a}}{(\veps_a - \veps_n)^2} \matrixel{a}{U}{a}
\,.
\end{align}
The summations marked by the prime here run over the complete spectrum excluding the equal-energy states that yield the vanishing denominators.

The key advantages of the recursive representation over the standard formulas are the universality and the computational efficiency. This method was used in Ref.~\cite{rozenbaum:14:pra} to find the higher-order contributions to the Zeeman and Stark shifts in hydrogenlike atoms, and the perfect agreement with the all-order A-DKB method was demonstrated. In Ref.~\cite{glazov:17:nimb} this method has been generalized to evaluate the higher-order contributions of the interelectronic interaction in few-electron ions.

Obviously, the $k$th-order correction $\veps_a^{(k)}$ can be found as the derivative of $\tveps_a$ with respect to $\lambda$:
\begin{equation}
\label{eq:ek}
  \veps_a^{(k)} = \left. \frac{1}{k!}\,\frac{d^k}{{d \lambda^k}} \, \tveps_a \right| _ {\lambda =0} 
 \,.
\end{equation}
Below we employ the generalization of this formula to the matrix elements of some operator in order to find the one-photon-exchange correction to $\gk{k}{}$.
%
\subsection{One-photon-exchange correction}
%
\begin{figure}
  \includegraphics{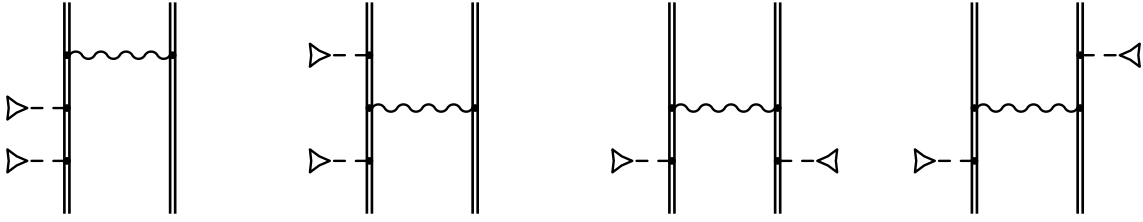}
  \caption{One-photon-exchange Feynman diagrams for the second-order Zeeman effect. The double line indicates the electron propagator in the binding potential, the wavy line indicates the photon propagator, and the dashed line terminated with the triangle denotes the interaction with the magnetic field.
  \label{fig:g-2-1ph}}
\end{figure}
\begin{figure}
  \includegraphics{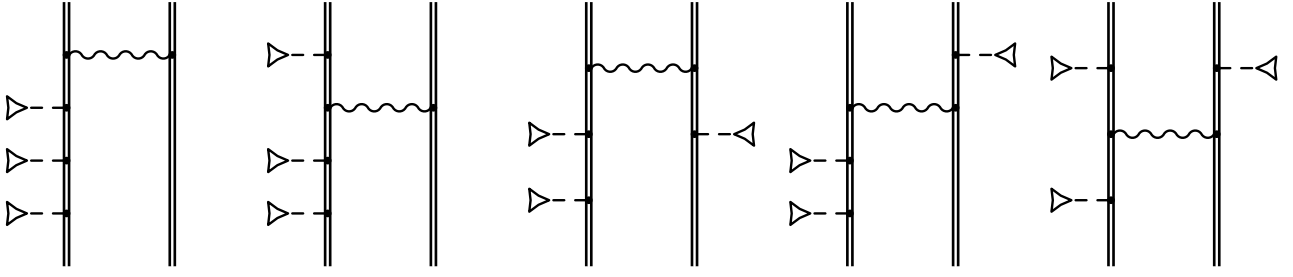}
  \caption{One-photon-exchange Feynman diagrams for the third-order Zeeman effect. The notations are the same as in Fig.~\ref{fig:g-2-1ph}.
  \label{fig:g-3-1ph}}
\end{figure}
The results for the third-order Zeeman effect in boronlike ions obtained in Ref.~\cite{varentsova:17:nimb} with Coulomb and two different (core-Hartree and Kohn-Sham) screening potentials demonstrated the significance of the interelectronic interaction. For this reason, the rigorous evaluation of this effect beyond the screening-potential approximation is in demand. The interelectronic-interaction correction of the first order in $1/Z$ is represented by the one-photon-exchange diagrams. In Figs.~\ref{fig:g-2-1ph} and~\ref{fig:g-3-1ph} these diagrams are presented for the second- and third-order effects, respectively.

The contribution of the one-photon exchange to the energy $\veps_a$ of the valence electron is given by the expression,
\begin{equation}
\label{eq:1ph}
  {\Delta \veps_a} =  \sum_{b} \left(\matrixelb{a\,b}{I(0)}{a\,b} - \matrixelb{a\,b}{I(\omega_{ab})}{b\,a} \right)
    - \matrixelb{a}{\Vscr}{a}
\,,
\end{equation}
where the summation runs over the core states ($1s$ and $2s$ with angular momentum projection $+1/2$ and $-1/2$), the electron-electron interaction operator $I$ in the Feynman gauge reads
\begin{equation}
\label{eq:I}
  I(\omega, \boldsymbol{r}_{12}) = \alpha\, (1-\balpha_1 \balpha_2)\, \frac{\exp(i |\omega| r_{12})}{r_{12}}
\,,
\end{equation}
and $\omega_{ab}=\veps_a-\veps_b$. The last term in Eq.~(\ref{eq:1ph}) is due to the inclusion of the screening potential into the Dirac equation~(\ref{eq:D}). In the presence of magnetic field the one-photon-exchange correction should be evaluated using the wave functions $\ket{\ta}$, $\ket{\tb}$ and the energies $\tveps_a$, $\tveps_b$,
\begin{equation}
\label{eq:1ph-t}
  {\Delta \tveps_a} =  \sum_{\tb} \left(\matrixelb{\ta\,\tb}{I(0)}{\ta\,\tb} - \matrixelb{\ta\,\tb}{I(\tomega_{ab})}{\tb\,\ta} \right) 
    - \matrixelb{\ta}{\Vscr}{\ta}
\,,
\end{equation}
where $\tomega_{ab} = \tveps_a - \tveps_b$. As well as $\tveps_a$ and $\tveps_b$, the correction $\Delta \tveps_a$ is the function of $\lambda$ and includes the contributions of all orders in magnetic field. Now the corresponding correction to the Zeeman effect of the order $k$ can be found by differentiating $\Delta \tveps_a$ with respect to $\lambda$,
\begin{equation}
\label{eq:delta-gk}
  \dgkinto{k} \equiv \Delta \veps_a^{(k)} = \left. \frac{1}{k!}  \frac{d^k {\Delta \tveps_a}}{{d \lambda^k}} \right|_{\lambda =0}
\,.
\end{equation}
Explicit evaluation of this expression for $k=1$ leads to the formula for the one-photon-exchange correction to the $g$ factor, presented, e.g., in Ref.~\cite{shabaev:02:pra}, where it has been calculated for lithiumlike ions. The corresponding correction for the second-order Zeeman effect in boronlike ions has been calculated in Ref.~\cite{agababaev:17:nimb} according to an explicit formula (not presented in that work), which is quite similar to the one obtained for the nuclear magnetic shielding \cite{moskovkin:08:os,moskovkin:08:pra}. The formula for $\dgkinto{3}$ is even more lengthy. As an example, we present the irreducible part of the first diagram in Fig.~\ref{fig:g-3-1ph},
\begin{align}
  \dgk{3}{\mathrm{int,1-A}} = 
    &\,2\, {\sum_{n_1,n_2,n_3}}' \, \frac{\matrixelb{a}{U}{n_1}\matrixelb{n_1}{U}{n_2}\matrixelb{n_2}{U}{n_3}\matrixelb{n_3\,b}{I(0)}{a\,b}}{(\veps_a-\veps_{n_1})(\veps_a-\veps_{n_2})(\veps_a-\veps_{n_3})}
\nonumber\\
  + &\,2\, {\sum_{n_1,n_2,n_3}}' \, \frac{\matrixelb{b}{U}{n_1}\matrixelb{n_1}{U}{n_2}\matrixelb{n_2}{U}{n_3}\matrixelb{a\,n_3}{I(0)}{a\,b}}{(\veps_b-\veps_{n_1})(\veps_b-\veps_{n_2})(\veps_b-\veps_{n_3})}
\nonumber\\
  - &\,2\, {\sum_{n_1,n_2,n_3}}' \, \frac{\matrixelb{a}{U}{n_1}\matrixelb{n_1}{U}{n_2}\matrixelb{n_2}{U}{n_3}\matrixelb{n_3\,b}{I(\omega_{ab})}{b\,a}}{(\veps_a-\veps_{n_1})(\veps_a-\veps_{n_2})(\veps_a-\veps_{n_3})}
\nonumber\\
  - &\,2\, {\sum_{n_1,n_2,n_3}}' \, \frac{\matrixelb{b}{U}{n_1}\matrixelb{n_1}{U}{n_2}\matrixelb{n_2}{U}{n_3}\matrixelb{a\,n_3}{I(\omega_{ab})}{b\,a}}{(\veps_b-\veps_{n_1})(\veps_b-\veps_{n_2})(\veps_b-\veps_{n_3})}
\,.
\end{align}
We mention that, apart from the irreducible contributions of each diagram in Fig.~\ref{fig:g-3-1ph}, there exist also numerous reducible contributions, including the terms with the derivatives of $I(\omega)$ up to the third order. The straightforward calculations according to these formulas are rather involved and time-consuming. In this work, instead of employing these formulas, we evaluate the one-photon-exchange correction to the second- and third-order Zeeman effect according to Eq.~(\ref{eq:delta-gk}). Below we discuss some details of the calculation.
%
\subsection{Numerical implementation}
\label{sec:num}
%
As a first step, the finite basis set of the solutions to the Dirac equation~(\ref{eq:D}) without magnetic field is constructed in the framework of the DKB method \cite{shabaev:04:prl} based on the B-splines \cite{sapirstein:96:jpb}. Then within the iterative procedure [Eqs.~(\ref{eq:rec-en}), (\ref{eq:rec-na}), and (\ref{eq:rec-aa})] the coefficients $\veps_a^{(k)}$ and $\braket{n}{a^{(k)}}$ are calculated for the $1s$, $2s$, and $2p_j$ states up to the third order ($k=1,2,3$). With these coefficients the energies $\tveps_a$, $\tveps_b$ and the wave functions $\ket{\ta}$, $\ket{\tb}$ can be calculated for arbitrary values of $\lambda$ according to Eqs.~(\ref{eq:ten}) and (\ref{eq:ta}). Since the magnetic-interaction operator $\Vmagn$ breaks the spherical symmetry of the Hamiltonian, the wave function $\ket{\ta}$ is no longer an eigenfunction of the total angular momentum operator. Nevertheless, it can be represented as a sum $\ket{\ta} = \sum_{\kappa} \ket{\ta_\kappa}$, where $\ket{\ta_\kappa}$ is the partial-wave contribution of definite value of the angular momentum-parity quantum number $\kappa=(j+1/2)(-1)^{j+l+1/2}$. It is convenient to treat these contributions separately in order to use the standard methods of angular integration for the matrix elements. The summations over $k$ and $n$ in Eq.~(\ref{eq:ta}) are implemented numerically for each partial-wave component $\ket{\ta_\kappa}$. Since the operator $\Vmagn$ mixes only the states with $\kappa_2 = \kappa_1$ and $\kappa_2 = -\kappa_1 \pm 1$, there are in general not more than 3, 5, and 7 nonzero contributions $\ket{\ta_\kappa}$ with different values of $\kappa$ in the first, second, and third order of the perturbation theory in $\Vmagn$, respectively.

With the energies $\tveps_a$, $\tveps_b$ and the wave functions $\ket{\ta}$, $\ket{\tb}$, the one-photon-exchange correction $\Delta \tveps_a$ can be calculated according to Eq.~(\ref{eq:1ph-t}) for arbitrary values of $\lambda$. The standard methods of calculation can be employed owing to the partial-wave expansion of the wave functions $\ket{\ta}$, $\ket{\tb}$ and of the interaction operator $I(\omega,\bfr_{12})$,
\begin{equation}
\label{eq:I-lm}
  I(\omega, \bfr_{12}) = \alpha\, (1-\balpha_1 \balpha_2)\, \sum_{L=0}^{\infty}\, g_L(\omega,r_1,r_2) \, \frac{4\pi}{2L+1} \, \sum_{M=-L}^{L} Y_{LM}(\hbfr_1) Y^*_{LM}(\hbfr_2)
\,,
\end{equation}
where the explicit formulas for $g_L$ can be found, e.g., in Ref.~\cite{varshalovich}. Since we are interested in the lowest-order terms ($k\le 3$) of the perturbation theory, the partial-wave summations are finite due to the well-known selection rules. The contribution of the one-photon exchange to the Zeeman effect of the order $k$ is found by differentiation with respect to $\lambda$ employing the finite difference method. The symmetric equidistant sets of points $\lambda$ = $-n\lambda_0$, ..., $-\lambda_0$, $0$, $\lambda_0$, ..., $n\lambda_0$ with $n =3, 4$ are used. Values of $\lambda_0$ are chosen so as to achieve stable results with respect to variation of $n$ and $\lambda_0$. Optimal values of $\lambda_0$ depend on the nuclear charge $Z$, the state $\ket{a}$, and the order $k$ of the perturbation theory.
%
\section{Results and discussion}
%
We have calculated the leading contribution and the one-photon-exchange correction to the second- and third-order Zeeman effect for the $\stJ$ states in boronlike ions in the range $Z=6$--$92$. The contributions of the zeroth and first order in magnetic field are also obtained in the course of calculation and used to control the numerical accuracy. In order to account partly for the higher-order contributions of the interelectronic interaction and to estimate the corresponding uncertainty, we use five different screening potentials: core-Hartree, Kohn-Sham, Dirac-Slater, local Dirac-Fock, and Perdew-Zunger. The description of these potentials can be found, e.g., in Refs.~\cite{kohn:65:pr,perdew:81:prb,sapirstein:02:pra,shabaev:05:pra}. The relativistic approach employed is valid to all orders in $\aZ$. In particular, the contribution of the negative-energy states, which is very important for the Zeeman effect, is taken into account.
%
\subsection{Quadratic Zeeman effect}
%
The total value of the second-order coefficient including the interelectronic-interaction correction is denoted as $\gBBint$,
\begin{equation}
  \gBBint = \gBB{0} + \dgBBinto + \dots
\,,
\end{equation}
where $\gBB{0}$ is the leading-order contribution and $\dgBBinto$ is the one-photon-exchange correction. For evaluation of $\gBB{0}$ and $\dgBBinto$ we restrict the summations in Eqs.~(\ref{eq:ten}) and (\ref{eq:ta}) to the values $k \le 2$. The leading-order contribution $\gBB{0}$ is calculated directly by the recursive equations~(\ref{eq:rec-en}), (\ref{eq:rec-na}), and (\ref{eq:rec-aa}). The one-photon-exchange correction $\dgBBinto$ is obtained by numerical differentiation of $\Delta \tveps_a$ with respect to $\lambda$ as described in Section~\ref{sec:num}. In addition, we evaluate $\dgBBinto$ in a straightforward manner by the explicit formulas, as it was done in Ref.~\cite{agababaev:17:nimb} for boronlike argon. The values obtained in both ways are in perfect agreement.

The results for the $\stgr$ state of boronlike ions in the range $Z=6$--$92$ are given in Table \ref{tab:g2_2p_12_12}. The values of $\gBB{0}$, $\dgBBinto$, and their sum $\gBBint$ are presented for the different screening potentials. The spread of the results can serve as an estimation of the unknown higher-order contributions of the interelectronic interaction. One can see that the spread for $\gBBint$ is significantly smaller than for $\gBB{0}$. This means that the first-order correction $\dgBBinto$ accounts for the dominant part of the interelectronic-interaction effect. The corresponding results for the $\stex$ state are given in Tables \ref{tab:g2_2p_32_12} and \ref{tab:g2_2p_32_32} for $M_J=\pm 1/2$ and $M_J=\pm 3/2$, respectively.

In Ref.~\cite{agababaev:17:nimb} the one-loop QED correction to the second-order Zeeman effect has been calculated for boronlike argon, in addition to the one-photon exchange. So, the most accurate to-date value of $\gBB{}$ for $Z=18$ is presented in that paper, while the systematic calculations of the QED correction for a wide range of $Z$ are presently underway.
%
\subsection{Cubic Zeeman effect}
%
By analogy with the second order, we evaluate the third-order coefficient $\gBBBint$ as the sum of the leading-order contribution $\gBBB{0}$ and the one-photon-exchange correction $\dgBBBinto$,
\begin{equation}
  \gBBBint = \gBBB{0} + \dgBBBinto + \dots
\,.
\end{equation}
In this case, the summations in Eqs.~(\ref{eq:ten}) and (\ref{eq:ta}) are restricted to the values $k \le 3$. The leading term $\gBBB{0}$ is evaluated within the recursive scheme, according to Eqs.~(\ref{eq:rec-en}), (\ref{eq:rec-na}), and (\ref{eq:rec-aa}), while the contribution of the one-photon exchange $\dgBBBinto$ is obtained by numerical differentiation of $\Delta \tveps_a$ with respect to $\lambda$. We stress that the straightforward evaluation of $\dgBBBinto$ within the perturbation theory would be rather involved and time-consuming.

In Tables \ref{tab:g3_2p_12_12}, \ref{tab:g3_2p_32_12}, and \ref{tab:g3_2p_32_32} we present the results for the state $\stgr$ with $M_J=1/2$ and for the state $\stex$ with $M_J=1/2$ and $M_J=3/2$, respectively. The corresponding results for the negative values of $M_J$ are immediately obtained by the symmetry relation: $\gBBB{}(-M_J)=-\gBBB{}(M_J)$. The higher-order contributions of the interelectronic interaction can be estimated from the spread of the results obtained with the different screening potentials. One can see that the spread is generally much smaller for $\gBBBint$ than for $\gBBB{0}$, so the prevailing contribution of the interelectronic interaction is taken into account.

The third-order Zeeman effect can be considered as a correction to the $g$ factor scaling as $B^2$,
\begin{equation}
\label{eq:g-g-3}
  \delta g = \lambda^2 \gBBB{}(M_J) / M_J
\,.
\end{equation}
In Table~\ref{tab:delta_g3} we present the results for $\delta g$ at the field of 1 Tesla based on the value of $\gBBBint$ obtained with the local Dirac-Fock potential. In order to find $\delta g$ at the field of $\beta$ Tesla the value from the Table~\ref{tab:delta_g3} has to be multiplied by $\beta^2$. For example, in boronlike argon at the field of 7 Tesla (as implied by the ARTEMIS experimental setup) we find $\delta g = 2.9 \times 10^{-9}$ for the ground state and $\delta g = -2.9 \times 10^{-9}$ for the $\stex$ state with $M_J=\pm 1/2$. The third-order shift of the levels with $M_J=\pm 3/2$ is six orders of magnitude smaller. For boronlike lead this correction is in the range $10^{-16}$--$10^{-14}$ assuming the magnetic field of several Tesla.
%
\section{Conclusion}
%
Relativistic calculations of the quadratic and cubic Zeeman effect have been performed for the ground $\lstgr$ and the first excited $\lstex$ states of boronlike ions in the range $Z=6$--$92$. The one-photon-exchange correction has been evaluated to all orders in $\aZ$. The higher-order interelectronic-interaction contributions have been taken into account approximately by the use of the effective screening potentials. The cubic Zeeman effect has been presented in terms of the field-dependent $g$-factor correction. Its value at the field of 1 Tesla ranges from $10^{-18}$ to almost $10^{-5}$ for the ions under consideration. These results are important for interpretation of experimental data, in particular, for the $g$-factor measurements in boronlike argon in the ARTEMIS experiment at GSI.
%
\section*{Acknowledgements}
%
We thank Aleksei Malyshev, Wolfgang Quint, and Manuel Vogel for valuable discussions. The work was supported by RFBR (Grant No.~16-02-00334), by SPbSU-DFG (Grants No. 11.65.41.2017 and No.~STO~346/5-1), by SPbSU (Grants No.~11.40.538.2017 and 11.42.688.2017), by DFG (Grant No. VO 1707/1-3), and by the FAIR--Russia Research Center (FRRC). V.A.A. acknowledges the support by the German--Russian Interdisciplinary Science Center (G-RISC). The numerical computations were performed at the St. Petersburg State University Computing Center.
%
%

%
\newpage
\begin{longtable}{rlr@{}lr@{}lr@{}lr@{}lr@{}l}
\caption{\label{tab:g2_2p_12_12}
Second-order Zeeman effect for the ground state ($\stgr$, $M_J=\pm 1/2$) of boronlike ions in terms of the dimensionless coefficient $\gBB{}$, see Eq.~(\ref{eq:g-2}). The leading-order term $\gBB{0}$, the one-photon-exchange correction $\dgBBinto$, and their sum $\gBBint$ are given for different screening potentials: core-Hartree (CH), Kohn-Sham (KS), Dirac-Slater (DS), local Dirac-Fock (LDF), and Perdew-Zunger (PZ).
} 
\vspace{1cm}
\\
		\hline
		\hline
	      $Z$
	& 
	& \multicolumn{2}{c}{CH}
	& \multicolumn{2}{c}{KS}
	& \multicolumn{2}{c}{DS} 
	& \multicolumn{2}{c}{LDF}
	& \multicolumn{2}{c}{PZ} \\
		\hline
	& $\gBB{0}$    &  $-$1&.1733$\dexp{7}$ & $-$1&.1155$\dexp{7}$	& $-$0&.8269$\dexp{7}$ & $-$1&.2454$\dexp{7}$ & $-$0&.8866$\dexp{7}$ \\
6   & $\dgBBinto$  &  $-$0&.0674$\dexp{7}$ & $-$0&.0638$\dexp{7}$	& $-$0&.4297$\dexp{7}$ & $-$0&.0114$\dexp{7}$ & $-$0&.3804$\dexp{7}$ \\
	& $\gBBint$    &  $-$1&.2407$\dexp{7}$ & $-$1&.1794$\dexp{7}$	& $-$1&.2567$\dexp{7}$ & $-$1&.2568$\dexp{7}$ & $-$1&.2671$\dexp{7}$ \\
		\hline
	& $\gBB{0}$    &  $-$2&.0285$\dexp{6}$ & $-$1&.9533$\dexp{6}$	& $-$1&.6204$\dexp{6}$ & $-$2&.1108$\dexp{6}$ & $-$1&.7037$\dexp{6}$ \\
8   & $\dgBBinto$  &  $-$0&.1564$\dexp{6}$ & $-$0&.1982$\dexp{6}$	& $-$0&.5534$\dexp{6}$ & $-$0&.0836$\dexp{6}$ & $-$0&.4835$\dexp{6}$ \\
	& $\gBBint$    &  $-$2&.1849$\dexp{6}$ & $-$2&.1516$\dexp{6}$	& $-$2&.1738$\dexp{6}$ & $-$2&.1945$\dexp{6}$ & $-$2&.1872$\dexp{6}$ \\
		\hline
	& $\gBB{0}$    &  $-$6&.1710$\dexp{5}$ & $-$5&.9872$\dexp{5}$	& $-$5&.2180$\dexp{5}$ & $-$6&.3390$\dexp{5}$ & $-$5&.4245$\dexp{5}$ \\
10  & $\dgBBinto$  &  $-$0&.4190$\dexp{5}$ & $-$0&.5534$\dexp{5}$	& $-$1&.3381$\dexp{5}$ & $-$0&.2678$\dexp{5}$ & $-$1&.1589$\dexp{5}$ \\
	& $\gBBint$    &  $-$6&.5901$\dexp{5}$ & $-$6&.5406$\dexp{5}$	& $-$6&.5562$\dexp{5}$ & $-$6&.6069$\dexp{5}$ & $-$6&.5835$\dexp{5}$ \\
		\hline
	& $\gBB{0}$    &  $-$2&.4823$\dexp{5}$ & $-$2&.4203$\dexp{5}$	& $-$2&.1700$\dexp{5}$ & $-$2&.5310$\dexp{5}$ & $-$2&.2397$\dexp{5}$ \\
12  & $\dgBBinto$  &  $-$0&.1452$\dexp{5}$ & $-$0&.1954$\dexp{5}$	& $-$0&.4467$\dexp{5}$ & $-$0&.1010$\dexp{5}$ & $-$0&.3847$\dexp{5}$ \\
	& $\gBBint$    &  $-$2&.6276$\dexp{5}$ & $-$2&.6157$\dexp{5}$	& $-$2&.6167$\dexp{5}$ & $-$2&.6321$\dexp{5}$ & $-$2&.6244$\dexp{5}$ \\
		\hline
	& $\gBB{0}$    &  $-$1&.1834$\dexp{5}$ & $-$1&.1580$\dexp{5}$	& $-$1&.0576$\dexp{5}$ & $-$1&.2011$\dexp{5}$ & $-$1&.0861$\dexp{5}$ \\
14  & $\dgBBinto$  &  $-$0&.0602$\dexp{5}$ & $-$0&.0819$\dexp{5}$	& $-$0&.1820$\dexp{5}$ & $-$0&.0441$\dexp{5}$ & $-$0&.1562$\dexp{5}$ \\
	& $\gBBint$    &  $-$1&.2437$\dexp{5}$ & $-$1&.2400$\dexp{5}$	& $-$1&.2396$\dexp{5}$ & $-$1&.2452$\dexp{5}$ & $-$1&.2424$\dexp{5}$ \\
		\hline
	& $\gBB{0}$    &  $-$6&.3296$\dexp{4}$ & $-$6&.2095$\dexp{4}$	& $-$5&.7453$\dexp{4}$ & $-$6&.4042$\dexp{4}$ & $-$5&.8795$\dexp{4}$ \\
16  & $\dgBBinto$  &  $-$0&.2839$\dexp{4}$ & $-$0&.3897$\dexp{4}$	& $-$0&.8509$\dexp{4}$ & $-$0&.2155$\dexp{4}$ & $-$0&.7279$\dexp{4}$ \\
	& $\gBBint$    &  $-$6&.6135$\dexp{4}$ & $-$6&.5992$\dexp{4}$	& $-$6&.5963$\dexp{4}$ & $-$6&.6197$\dexp{4}$ & $-$6&.6075$\dexp{4}$ \\
		\hline
	& $\gBB{0}$    &  $-$3&.6782$\dexp{4}$ & $-$3&.6157$\dexp{4}$	& $-$3&.3779$\dexp{4}$ & $-$3&.7135$\dexp{4}$ & $-$3&.4474$\dexp{4}$ \\
18  & $\dgBBinto$  &  $-$0&.1470$\dexp{4}$ & $-$0&.2033$\dexp{4}$	& $-$0&.4393$\dexp{4}$ & $-$0&.1146$\dexp{4}$ & $-$0&.3750$\dexp{4}$ \\
	& $\gBBint$    &  $-$3&.8253$\dexp{4}$ & $-$3&.8191$\dexp{4}$	& $-$3&.8172$\dexp{4}$ & $-$3&.8281$\dexp{4}$ & $-$3&.8224$\dexp{4}$ \\
		\hline
	& $\gBB{0}$    &  $-$2&.2760$\dexp{4}$ & $-$2&.2409$\dexp{4}$	& $-$2&.1091$\dexp{4}$ & $-$2&.2941$\dexp{4}$ & $-$2&.1479$\dexp{4}$ \\
20  & $\dgBBinto$  &  $-$0&.0819$\dexp{4}$ & $-$0&.1140$\dexp{4}$	& $-$0&.2446$\dexp{4}$ & $-$0&.0652$\dexp{4}$ & $-$0&.2084$\dexp{4}$ \\
	& $\gBBint$    &  $-$2&.3579$\dexp{4}$ & $-$2&.3549$\dexp{4}$	& $-$2&.3538$\dexp{4}$ & $-$2&.3593$\dexp{4}$ & $-$2&.3564$\dexp{4}$ \\
		\hline
	& $\gBB{0}$    &  $-$1&.4792$\dexp{4}$ & $-$1&.4583$\dexp{4}$	& $-$1&.3807$\dexp{4}$ & $-$1&.4892$\dexp{4}$ & $-$1&.4037$\dexp{4}$ \\
22  & $\dgBBinto$  &  $-$0&.0483$\dexp{4}$ & $-$0&.0677$\dexp{4}$	& $-$0&.1446$\dexp{4}$ & $-$0&.0391$\dexp{4}$ & $-$0&.1230$\dexp{4}$ \\
	& $\gBBint$    &  $-$1&.5276$\dexp{4}$ & $-$1&.5260$\dexp{4}$	& $-$1&.5253$\dexp{4}$ & $-$1&.5283$\dexp{4}$ & $-$1&.5267$\dexp{4}$ \\
		\hline
	& $\gBB{0}$    &  $-$1&.0000$\dexp{4}$ & $-$0&.9869$\dexp{4}$ & $-$0&.9389$\dexp{4}$ & $-$1&.0058$\dexp{4}$ & $-$0&.9532$\dexp{4}$ \\
24  & $\dgBBinto$  &  $-$0&.0298$\dexp{4}$ & $-$0&.0421$\dexp{4}$	& $-$0&.0897$\dexp{4}$ & $-$0&.0245$\dexp{4}$ & $-$0&.0762$\dexp{4}$ \\
	& $\gBBint$    &  $-$1&.0299$\dexp{4}$ & $-$1&.0291$\dexp{4}$	& $-$1&.0286$\dexp{4}$ & $-$1&.0304$\dexp{4}$ & $-$1&.0294$\dexp{4}$ \\
		\hline
	& $\gBB{0}$    &  $-$2&.7502$\dexp{3}$ & $-$2&.7216$\dexp{3}$	& $-$2&.6220$\dexp{3}$ & $-$2&.7594$\dexp{3}$ & $-$2&.6521$\dexp{3}$ \\
32  & $\dgBBinto$  &  $-$0&.0602$\dexp{3}$ & $-$0&.0877$\dexp{3}$	& $-$0&.1864$\dexp{3}$ & $-$0&.0518$\dexp{3}$ & $-$0&.1576$\dexp{3}$ \\
	& $\gBBint$    &  $-$2&.8105$\dexp{3}$ & $-$2&.8093$\dexp{3}$	& $-$2&.8084$\dexp{3}$ & $-$2&.8113$\dexp{3}$ & $-$2&.8098$\dexp{3}$ \\
		\hline
	& $\gBB{0}$    &  $-$2&.4067$\dexp{2}$ & $-$2&.3877$\dexp{2}$	& $-$2&.3320$\dexp{2}$ & $-$2&.4076$\dexp{2}$ & $-$2&.3492$\dexp{2}$ \\
54  & $\dgBBinto$  &  $-$0&.0252$\dexp{2}$ & $-$0&.0439$\dexp{2}$ & $-$0&.0992$\dexp{2}$ & $-$0&.0247$\dexp{2}$ & $-$0&.0826$\dexp{2}$ \\
	& $\gBBint$    &  $-$2&.4319$\dexp{2}$ & $-$2&.4316$\dexp{2}$ & $-$2&.4313$\dexp{2}$ & $-$2&.4324$\dexp{2}$ & $-$2&.4318$\dexp{2}$ \\
		\hline
	& $\gBB{0}$    &  $-$2&.3238$\dexp{}$  & $-$2&.3028$\dexp{}$  & $-$2&.2576$\dexp{}$  & $-$2&.3176$\dexp{}$  & $-$2&.2721$\dexp{}$ \\
82  & $\dgBBinto$  &     0&.0087$\dexp{}$  & $-$0&.0120$\dexp{}$  & $-$0&.0570$\dexp{}$  & $ $0&.0020$\dexp{}$  & $-$0&.0429$\dexp{}$ \\
	& $\gBBint$    &  $-$2&.3150$\dexp{}$  & $-$2&.3148$\dexp{}$  & $-$2&.3146$\dexp{}$  & $-$2&.3155$\dexp{}$  & $-$2&.3151$\dexp{}$ \\
		\hline
	& $\gBB{0}$    &  $-$2&.1403$\dexp{}$  & $-$2&.1207$\dexp{}$	& $-$2&.0790$\dexp{}$  & $-$2&.1343$\dexp{}$  & $-$2&.0924$\dexp{}$ \\
83  & $\dgBBinto$  &     0&.0099$\dexp{}$  & $-$0&.0095$\dexp{}$	& $-$0&.0510$\dexp{}$  & $ $0&.0034$\dexp{}$  & $-$0&.0380$\dexp{}$ \\
	& $\gBBint$    &  $-$2&.1304$\dexp{}$  & $-$2&.1302$\dexp{}$	& $-$2&.1301$\dexp{}$  & $-$2&.1309$\dexp{}$  & $-$2&.1305$\dexp{}$ \\
		\hline
	& $\gBB{0}$    &  $-$9&.9072		   & $-$9&.7978 		& $-$9&.5950 		 & $-$9&.8619         & $-$9&.6626        \\
92  & $\dgBBinto$  &     0&.1539		   &    0&.0455 		& $-$0&.1564 		 & $ $0&.1059         & $-$0&.0915        \\
	& $\gBBint$    &  $-$9&.7533		   & $-$9&.7522 		& $-$9&.7514 		 & $-$9&.7559         & $-$9&.7542        \\
		\hline
		\hline
\end{longtable}
\newpage
\begin{longtable}{rlr@{}lr@{}lr@{}lr@{}lr@{}l}
\caption{\label{tab:g2_2p_32_12}
Second-order Zeeman effect for the first excited state ($\stex$, $M_J=\pm 1/2$) of boronlike ions in terms of the dimensionless coefficient $\gBB{}$, see Eq.~(\ref{eq:g-2}). The leading-order term $\gBB{0}$, the one-photon-exchange correction $\dgBBinto$, and their sum $\gBBint$ are given for different screening potentials: core-Hartree (CH), Kohn-Sham (KS), Dirac-Slater (DS), local Dirac-Fock (LDF), and Perdew-Zunger (PZ).
}
\vspace{1cm}
\\
		\hline
		\hline
		$Z$
		& 
		& \multicolumn{2}{c}{CH}
		& \multicolumn{2}{c}{KS}
		& \multicolumn{2}{c}{DS} 
		& \multicolumn{2}{c}{LDF}
		& \multicolumn{2}{c}{PZ} \\
		\hline
		& $\gBB{0}$    &  $ $1&.1767$\dexp{7}$ & $ $1&.1190$\dexp{7}$	& $ $0&.8298$\dexp{7}$  & $ $1&.2488$\dexp{7}$ & $ $0&.8896$\dexp{7}$\\
	6   & $\dgBBinto$  &  $ $0&.0683$\dexp{7}$ & $ $0&.0646$\dexp{7}$	& $ $0&.4313$\dexp{7}$  & $ $0&.0122$\dexp{7}$ & $ $0&.3818$\dexp{7}$\\
		& $\gBBint$    &  $ $1&.2450$\dexp{7}$ & $ $1&.1837$\dexp{7}$	& $ $1&.2611$\dexp{7}$  & $ $1&.2610$\dexp{7}$ & $ $1&.2714$\dexp{7}$\\
		\hline
		& $\gBB{0}$    &  $ $2&.0405$\dexp{6}$ & $ $1&.9656$\dexp{6}$	& $ $1&.6312$\dexp{6}$  & $ $2&.1231$\dexp{6}$ & $ $1&.7148$\dexp{6}$ \\
	8   & $\dgBBinto$  &  $ $0&.1592$\dexp{6}$ & $ $0&.2009$\dexp{6}$	& $ $0&.5574$\dexp{6}$  & $ $0&.0860$\dexp{6}$ & $ $0&.4870$\dexp{6}$ \\
		& $\gBBint$    &  $ $2&.1998$\dexp{6}$ & $ $2&.1666$\dexp{6}$	& $ $2&.1887$\dexp{6}$  & $ $2&.2092$\dexp{6}$ & $ $2&.2019$\dexp{6}$ \\
		\hline
		& $\gBB{0}$    &  $ $6&.2334$\dexp{5}$ & $ $6&.0504$\dexp{5}$	& $ $5&.2758$\dexp{5}$  & $ $6&.4026$\dexp{5}$ & $ $5&.4835$\dexp{5}$ \\
	10  & $\dgBBinto$  &  $ $0&.4307$\dexp{5}$ & $ $0&.5647$\dexp{5}$	& $ $1&.3545$\dexp{5}$  & $ $0&.2779$\dexp{5}$ & $ $1&.1734$\dexp{5}$ \\
		& $\gBBint$    &  $ $6&.6641$\dexp{5}$ & $ $6&.6151$\dexp{5}$	& $ $6&.6303$\dexp{5}$  & $ $6&.6805$\dexp{5}$ & $ $6&.6569$\dexp{5}$ \\
		\hline
		& $\gBB{0}$    &  $ $2&.5205$\dexp{5}$ & $ $2&.4589$\dexp{5}$	& $ $2&.2060$\dexp{5}$ & $ $2&.5698$\dexp{5}$  & $ $2&.2763$\dexp{5}$ \\
	12  & $\dgBBinto$  &  $ $0&.1511$\dexp{5}$ & $ $0&.2011$\dexp{5}$	& $ $0&.4548$\dexp{5}$ & $ $0&.1062$\dexp{5}$  & $ $0&.3919$\dexp{5}$ \\
		& $\gBBint$    &  $ $2&.6717$\dexp{5}$ & $ $2&.6600$\dexp{5}$	& $ $2&.6608$\dexp{5}$ & $ $2&.6760$\dexp{5}$  & $ $2&.6683$\dexp{5}$ \\
		\hline
		& $\gBB{0}$    &  $ $1&.2093$\dexp{5}$ & $ $1&.1840$\dexp{5}$	& $ $1&.0821$\dexp{5}$ & $ $1&.2273$\dexp{5}$  & $ $1&.1111$\dexp{5}$ \\
	14  & $\dgBBinto$  &  $ $0&.0636$\dexp{5}$ & $ $0&.0852$\dexp{5}$	& $ $0&.1867$\dexp{5}$ & $ $0&.0470$\dexp{5}$  & $ $0&.1603$\dexp{5}$ \\
		& $\gBBint$    &  $ $1&.2729$\dexp{5}$ & $ $1&.2693$\dexp{5}$	& $ $1&.2689$\dexp{5}$ & $ $1&.2744$\dexp{5}$  & $ $1&.2715$\dexp{5}$\\
		\hline
		& $\gBB{0}$    &  $ $6&.5160$\dexp{4}$ & $ $6&.3970$\dexp{4}$	& $ $5&.9237$\dexp{4}$ & $ $6&.5927$\dexp{4}$  & $ $6&.0603$\dexp{4}$ \\
	16  & $\dgBBinto$  &  $ $0&.3050$\dexp{4}$ & $ $0&.4101$\dexp{4}$	& $ $0&.8799$\dexp{4}$ & $ $0&.2340$\dexp{4}$  & $ $0&.7538$\dexp{4}$ \\
		& $\gBBint$    &  $ $6&.8210$\dexp{4}$ & $ $6&.8071$\dexp{4}$	& $ $6&.8037$\dexp{4}$ & $ $6&.8268$\dexp{4}$  & $ $6&.8142$\dexp{4}$ \\
		\hline
		& $\gBB{0}$    &  $ $3&.8189$\dexp{4}$ & $ $3&.7571$\dexp{4}$	& $ $3&.5133$\dexp{4}$ & $ $3&.8556$\dexp{4}$  & $ $3&.5844$\dexp{4}$\\
	18  & $\dgBBinto$  &  $ $0&.1611$\dexp{4}$ & $ $0&.2170$\dexp{4}$	& $ $0&.4586$\dexp{4}$ & $ $0&.1270$\dexp{4}$  & $ $0&.3922$\dexp{4}$\\
		& $\gBBint$    &  $ $3&.9800$\dexp{4}$ & $ $3&.9741$\dexp{4}$	& $ $3&.9719$\dexp{4}$ & $ $3&.9827$\dexp{4}$  & $ $3&.9767$\dexp{4}$\\
		\hline
		& $\gBB{0}$    &  $ $2&.3859$\dexp{4}$ & $ $2&.3512$\dexp{4}$	& $ $2&.2153$\dexp{4}$ & $ $2&.4050$\dexp{4}$  & $ $2&.2553$\dexp{4}$\\
	20  & $\dgBBinto$  &  $ $0&.0917$\dexp{4}$ & $ $0&.1236$\dexp{4}$	& $ $0&.2581$\dexp{4}$ & $ $0&.0739$\dexp{4}$  & $ $0&.2205$\dexp{4}$\\
		& $\gBBint$    &  $ $2&.4777$\dexp{4}$ & $ $2&.4748$\dexp{4}$	& $ $2&.4735$\dexp{4}$ & $ $2&.4790$\dexp{4}$  & $ $2&.4759$\dexp{4}$\\
		\hline
		& $\gBB{0}$    &  $ $1&.5674$\dexp{4}$ & $ $1&.5467$\dexp{4}$	& $ $1&.4662$\dexp{4}$ & $ $1&.5781$\dexp{4}$  & $ $1&.4900$\dexp{4}$\\
	22  & $\dgBBinto$  &  $ $0&.0554$\dexp{4}$ & $ $0&.0747$\dexp{4}$	& $ $0&.1544$\dexp{4}$ & $ $0&.0455$\dexp{4}$  & $ $0&.1318$\dexp{4}$\\
		& $\gBBint$    &  $ $1&.6229$\dexp{4}$ & $ $1&.6215$\dexp{4}$	& $ $1&.6206$\dexp{4}$ & $ $1&.6236$\dexp{4}$  & $ $1&.6219$\dexp{4}$\\
		\hline
		& $\gBB{0}$    &  $ $1&.0723$\dexp{4}$ & $ $1&.0594$\dexp{4}$ & $ $1&.0092$\dexp{4}$ & $ $1&.0786$\dexp{4}$  & $ $1&.0241$\dexp{4}$\\
	24  & $\dgBBinto$  &  $ $0&.0352$\dexp{4}$ & $ $0&.0473$\dexp{4}$	& $ $0&.0970$\dexp{4}$ & $ $0&.0293$\dexp{4}$  & $ $0&.0828$\dexp{4}$\\
		& $\gBBint$    &  $ $1&.1076$\dexp{4}$ & $ $1&.1068$\dexp{4}$	& $ $1&.1063$\dexp{4}$ & $ $1&.1079$\dexp{4}$  & $ $1&.1070$\dexp{4}$\\
		\hline
		& $\gBB{0}$    &  $ $3&.1282$\dexp{3}$ & $ $3&.1002$\dexp{3}$	& $ $2&.9921$\dexp{3}$ & $ $3&.1392$\dexp{3}$  & $ $3&.0248$\dexp{3}$\\
	32  & $\dgBBinto$  &  $ $0&.0812$\dexp{3}$ & $ $0&.1081$\dexp{3}$	& $ $0&.2152$\dexp{3}$ & $ $0&.0708$\dexp{3}$  & $ $0&.1835$\dexp{3}$\\
		& $\gBBint$    &  $ $3&.2094$\dexp{3}$ & $ $3&.2084$\dexp{3}$	& $ $3&.2074$\dexp{3}$ & $ $3&.2100$\dexp{3}$  & $ $3&.2084$\dexp{3}$\\
		\hline
		& $\gBB{0}$    &  $ $3&.5710$\dexp{2}$ & $ $3&.5525$\dexp{2}$	& $ $3&.4810$\dexp{2}$ & $ $3&.5747$\dexp{2}$  & $ $3&.5032$\dexp{2}$\\
	54  & $\dgBBinto$  &  $ $0&.0650$\dexp{2}$ & $ $0&.0834$\dexp{2}$ & $ $0&.1544$\dexp{2}$ & $ $0&.0616$\dexp{2}$  & $ $0&.1324$\dexp{2}$\\
		& $\gBBint$    &  $ $3&.6361$\dexp{2}$ & $ $3&.6360$\dexp{2}$ & $ $3&.6354$\dexp{2}$ & $ $3&.6363$\dexp{2}$  & $ $3&.6356$\dexp{2}$\\
		\hline
		& $\gBB{0}$    &  $ $6&.5589$\dexp{}$  & $ $6&.5354$\dexp{}$  & $ $6&.4485$\dexp{}$  & $ $6&.5561$\dexp{}$   & $ $6&.4766$\dexp{}$\\
	82  & $\dgBBinto$  &  $ $0&.0970$\dexp{}$  & $ $0&.1209$\dexp{}$  & $ $0&.2072$\dexp{}$  & $ $0&.1001$\dexp{}$   & $ $0&.1790$\dexp{}$\\
		& $\gBBint$    &  $ $6&.6560$\dexp{}$  & $ $6&.6563$\dexp{}$  & $ $6&.6557$\dexp{}$  & $ $6&.6562$\dexp{}$   & $ $6&.6556$\dexp{}$\\
		\hline
		& $\gBB{0}$    &  $ $6&.2428$\dexp{}$  & $ $6&.2205$\dexp{}$	& $ $6&.1388$\dexp{}$  & $ $6&.2398$\dexp{}$   & $ $6&.1653$\dexp{}$\\
	83  & $\dgBBinto$  &  $ $0&.0918$\dexp{}$  & $ $0&.1144$\dexp{}$	& $ $0&.1956$\dexp{}$  & $ $0&.0950$\dexp{}$   & $ $0&.1690$\dexp{}$\\
		& $\gBBint$    &  $ $6&.3347$\dexp{}$  & $ $6&.3350$\dexp{}$	& $ $6&.3344$\dexp{}$  & $ $6&.3349$\dexp{}$   & $ $6&.3344$\dexp{}$\\
		\hline
		& $\gBB{0}$    &  $ $4&.0903$\dexp{}$  & $ $4&.0762$\dexp{}$	&    4&.0273$\dexp{}$  & $ $4&.0866$\dexp{}$   & $ $4&.0436$\dexp{}$\\
	92  & $\dgBBinto$  &  $ $0&.0571$\dexp{}$  & $ $0&.0716$\dexp{}$	& $ $0&.1201$\dexp{}$  & $ $0&.0610$\dexp{}$   & $ $0&.1038$\dexp{}$\\
		& $\gBBint$    &  $ $4&.1475$\dexp{}$  & $ $4&.1478$\dexp{}$	& $ $4&.1474$\dexp{}$  & $ $4&.1476$\dexp{}$   & $ $4&.1474$\dexp{}$\\
		\hline
		\hline
\end{longtable}
\newpage
\begin{longtable}{rlr@{}lr@{}lr@{}lr@{}lr@{}l}
\caption{\label{tab:g2_2p_32_32}
Second-order Zeeman effect for the first excited state ($\stex$, $M_J=\pm 3/2$) of boronlike ions in terms of the dimensionless coefficient $\gBB{}$, see Eq.~(\ref{eq:g-2}). The leading-order term $\gBB{0}$, the one-photon-exchange correction $\dgBBinto$, and their sum $\gBBint$ are given for different screening potentials: core-Hartree (CH), Kohn-Sham (KS), Dirac-Slater (DS), local Dirac-Fock (LDF), and Perdew-Zunger (PZ).
}
\vspace{1cm}
    \\
		\hline
		\hline
		$Z$
		& 
		& \multicolumn{2}{c}{CH}
		& \multicolumn{2}{c}{KS}
		& \multicolumn{2}{c}{DS} 
		& \multicolumn{2}{c}{LDF}
		& \multicolumn{2}{c}{PZ} \\
		\hline
		& $\gBB{0}$    &  $ $2&.2595$\dexp{4}$ & $ $2&.6069$\dexp{4}$	& $ $2&.0518$\dexp{4}$ & $ $2&.3081$\dexp{4}$ & $ $1&.9417$\dexp{4}$ \\
	6   & $\dgBBinto$  &  $ $0&.3747$\dexp{4}$ & $-$0&.1263$\dexp{4}$	& $ $0&.7166$\dexp{4}$ & $ $0&.2930$\dexp{4}$ & $ $0&.7137$\dexp{4}$ \\
		& $\gBBint$    &  $ $2&.6343$\dexp{4}$ & $ $2&.4806$\dexp{4}$	& $ $2&.7684$\dexp{4}$ & $ $2&.6011$\dexp{4}$ & $ $2&.6555$\dexp{4}$ \\
		\hline
		& $\gBB{0}$    &  $ $8&.0271$\dexp{3}$ & $ $8&.6593$\dexp{3}$	&    7&.5425$\dexp{3}$ & $ $8&.2260$\dexp{3}$ & $ $7&.4066$\dexp{3}$ \\
	8   & $\dgBBinto$  &  $ $1&.2890$\dexp{3}$ & $ $0&.6016$\dexp{3}$	& $ $1&.9089$\dexp{3}$ & $ $1&.0162$\dexp{3}$ & $ $1&.8608$\dexp{3}$ \\
		& $\gBBint$    &  $ $9&.3162$\dexp{3}$ & $ $9&.2609$\dexp{3}$	& $ $9&.4514$\dexp{3}$ & $ $9&.2423$\dexp{3}$ & $ $9&.2675$\dexp{3}$ \\
		\hline
		& $\gBB{0}$    &  $ $4&.1600$\dexp{3}$ & $ $4&.3782$\dexp{3}$	&    3&.9688$\dexp{3}$ & $ $4&.2428$\dexp{3}$ & $ $3&.9337$\dexp{3}$ \\
	10  & $\dgBBinto$  &  $ $0&.5417$\dexp{3}$ & $ $0&.3205$\dexp{3}$	& $ $0&.7660$\dexp{3}$ & $ $0&.4375$\dexp{3}$ & $ $0&.7482$\dexp{3}$ \\
		& $\gBBint$    &  $ $4&.7018$\dexp{3}$ & $ $4&.6988$\dexp{3}$	& $ $4&.7349$\dexp{3}$ & $ $4&.6804$\dexp{3}$ & $ $4&.6819$\dexp{3}$ \\
		\hline
		& $\gBB{0}$    &  $ $2&.5516$\dexp{3}$ & $ $2&.6515$\dexp{3}$	& $ $2&.4564$\dexp{3}$ & $ $2&.5927$\dexp{3}$ & $ $2&.4437$\dexp{3}$\\
	12  & $\dgBBinto$  &  $ $0&.2747$\dexp{3}$ & $ $0&.1757$\dexp{3}$	& $ $0&.3817$\dexp{3}$ & $ $0&.2253$\dexp{3}$ & $ $0&.3738$\dexp{3}$\\
		& $\gBBint$    &  $ $2&.8263$\dexp{3}$ & $ $2&.8272$\dexp{3}$	& $ $2&.8381$\dexp{3}$ & $ $2&.8181$\dexp{3}$ & $ $2&.8175$\dexp{3}$\\
		\hline
		& $\gBB{0}$    &  $ $1&.7261$\dexp{3}$ & $ $1&.7799$\dexp{3}$	& $ $1&.6718$\dexp{3}$ & $ $1&.7492$\dexp{3}$ & $ $1&.6662$\dexp{3}$\\
	14  & $\dgBBinto$  &  $ $0&.1576$\dexp{3}$ & $ $0&.1047$\dexp{3}$	& $ $0&.2172$\dexp{3}$ & $ $0&.1307$\dexp{3}$ & $ $0&.2131$\dexp{3}$\\
		& $\gBBint$    &  $ $1&.8837$\dexp{3}$ & $ $1&.8847$\dexp{3}$	& $ $1&.8890$\dexp{3}$ & $ $1&.8799$\dexp{3}$ & $ $1&.8794$\dexp{3}$\\
		\hline
		& $\gBB{0}$    &  $ $1&.2455$\dexp{3}$ & $ $1&.2777$\dexp{3}$	& $ $1&.2116$\dexp{3}$ & $ $1&.2597$\dexp{3}$ & $ $1&.2088$\dexp{3}$\\
	16  & $\dgBBinto$  &  $ $0&.0986$\dexp{3}$ & $ $0&.0670$\dexp{3}$	& $ $0&.1352$\dexp{3}$ & $ $0&.0824$\dexp{3}$ & $ $0&.1329$\dexp{3}$\\
		& $\gBBint$    &  $ $1&.3441$\dexp{3}$ & $ $1&.3448$\dexp{3}$	& $ $1&.3468$\dexp{3}$ & $ $1&.3421$\dexp{3}$ & $ $1&.3417$\dexp{3}$\\
		\hline
		& $\gBB{0}$    &  $ $0&.9409$\dexp{3}$ & $ $0&.9617$\dexp{3}$	& $ $0&.9183$\dexp{3}$ & $ $0&.9502$\dexp{3}$ & $ $0&.9168$\dexp{3}$\\
	18  & $\dgBBinto$  &  $ $0&.0656$\dexp{3}$ & $ $0&.0453$\dexp{3}$	& $ $0&.0898$\dexp{3}$ & $ $0&.0552$\dexp{3}$ & $ $0&.0884$\dexp{3}$\\
		& $\gBBint$    &  $ $1&.0066$\dexp{3}$ & $ $1&.0071$\dexp{3}$	& $ $1&.0081$\dexp{3}$ & $ $1&.0055$\dexp{3}$ & $ $1&.0052$\dexp{3}$\\
		\hline
		& $\gBB{0}$    &  $ $7&.3570$\dexp{2}$ & $ $7&.4991$\dexp{2}$	& $ $7&.1986$\dexp{2}$ & $ $7&.4213$\dexp{2}$ & $ $7&.1897$\dexp{2}$\\
	20  & $\dgBBinto$  &  $ $0&.4590$\dexp{2}$ & $ $0&.3200$\dexp{2}$	& $ $0&.6265$\dexp{2}$ & $ $0&.3878$\dexp{2}$ & $ $0&.6176$\dexp{2}$\\
		& $\gBBint$    &  $ $7&.8161$\dexp{2}$ & $ $7&.8191$\dexp{2}$	& $ $7&.8251$\dexp{2}$ & $ $7&.8091$\dexp{2}$ & $ $7&.8073$\dexp{2}$\\
		\hline
		& $\gBB{0}$    &  $ $5&.9077$\dexp{2}$ & $ $6&.0090$\dexp{2}$	& $ $5&.7924$\dexp{2}$ & $ $5&.9538$\dexp{2}$ & $ $5&.7869$\dexp{2}$\\
	22  & $\dgBBinto$  &  $ $0&.3331$\dexp{2}$ & $ $0&.2339$\dexp{2}$	& $ $0&.4541$\dexp{2}$ & $ $0&.2825$\dexp{2}$ & $ $0&.4481$\dexp{2}$\\
		& $\gBBint$    &  $ $6&.2408$\dexp{2}$ & $ $6&.2429$\dexp{2}$	& $ $6&.2466$\dexp{2}$ & $ $6&.2363$\dexp{2}$ & $ $6&.2351$\dexp{2}$\\
		\hline
		& $\gBB{0}$    &  $ $4&.8462$\dexp{2}$ & $ $4&.9208$\dexp{2}$ & $ $4&.7597$\dexp{2}$ & $ $4&.8802$\dexp{2}$ & $ $4&.7561$\dexp{2}$\\
	24  & $\dgBBinto$  &  $ $0&.2491$\dexp{2}$ & $ $0&.1760$\dexp{2}$	& $ $0&.3396$\dexp{2}$ & $ $0&.2121$\dexp{2}$ & $ $0&.3353$\dexp{2}$\\
		& $\gBBint$    &  $ $5&.0954$\dexp{2}$ & $ $5&.0969$\dexp{2}$	& $ $5&.0993$\dexp{2}$ & $ $5&.0924$\dexp{2}$ & $ $5&.0915$\dexp{2}$\\
		\hline
		& $\gBB{0}$    &  $ $2&.5454$\dexp{2}$ & $ $2&.5734$\dexp{2}$	& $ $2&.5113$\dexp{2}$ & $ $2&.5581$\dexp{2}$ & $ $2&.5104$\dexp{2}$\\
	32  & $\dgBBinto$  &  $ $0&.0966$\dexp{2}$ & $ $0&.0691$\dexp{2}$	& $ $0&.1318$\dexp{2}$ & $ $0&.0831$\dexp{2}$ & $ $0&.1305$\dexp{2}$\\
		& $\gBBint$    &  $ $2&.6421$\dexp{2}$ & $ $2&.6425$\dexp{2}$	& $ $2&.6431$\dexp{2}$ & $ $2&.6412$\dexp{2}$ & $ $2&.6410$\dexp{2}$\\
		\hline
		& $\gBB{0}$    &  $ $8&.0025$\dexp{}$  & $ $8&.0524$\dexp{}$	& $ $7&.9360$\dexp{}$  & $ $8&.0228$\dexp{}$  & $ $7&.9354$\dexp{}$\\
	54  & $\dgBBinto$  &  $ $0&.1745$\dexp{}$  & $ $0&.1250$\dexp{}$  & $ $0&.2422$\dexp{}$  & $ $0&.1533$\dexp{}$  & $ $0&.2406$\dexp{}$\\
		& $\gBBint$    &  $ $8&.1770$\dexp{}$  & $ $8&.1775$\dexp{}$  & $ $8&.1783$\dexp{}$  & $ $8&.1762$\dexp{}$  & $ $8&.1760$\dexp{}$\\
		\hline
		& $\gBB{0}$    &  $ $3&.0688$\dexp{}$  & $ $3&.0815$\dexp{}$  & $ $3&.0499$\dexp{}$  & $ $3&.0723$\dexp{}$  & $ $3&.0496$\dexp{}$\\
	82  & $\dgBBinto$  &     0&.0406$\dexp{}$  &    0&.0279$\dexp{}$  & $ $0&.0597$\dexp{}$  & $ $0&.0369$\dexp{}$  & $ $0&.0596$\dexp{}$\\
		& $\gBBint$    &  $ $3&.1094$\dexp{}$  & $ $3&.1094$\dexp{}$  & $ $3&.1097$\dexp{}$  & $ $3&.1093$\dexp{}$  & $ $3&.1093$\dexp{}$\\
		\hline
		& $\gBB{0}$    &  $ $2&.9809$\dexp{}$  & $ $2&.9930$\dexp{}$	& $ $2&.9626$\dexp{}$  & $ $2&.9842$\dexp{}$  & $ $2&.9624$\dexp{}$\\
	83  & $\dgBBinto$  &     0&.0387$\dexp{}$  &    0&.0266$\dexp{}$	& $ $0&.0572$\dexp{}$  & $ $0&.0353$\dexp{}$  & $ $0&.0571$\dexp{}$\\
		& $\gBBint$    &  $ $3&.0196$\dexp{}$  & $ $3&.0197$\dexp{}$	& $ $3&.0199$\dexp{}$  & $ $3&.0195$\dexp{}$  & $ $3&.0196$\dexp{}$\\
		\hline
		& $\gBB{0}$    &  $ $2&.3182$\dexp{}$  & $ $2&.3268$\dexp{}$	& $ $2&.3047$\dexp{}$  & $ $2&.3199$\dexp{}$  & $ $2&.3045$\dexp{}$\\
	92  & $\dgBBinto$  &     0&.0256$\dexp{}$  &    0&.0171$\dexp{}$	& $ $0&.0393$\dexp{}$  & $ $0&.0238$\dexp{}$  & $ $0&.0393$\dexp{}$\\
		& $\gBBint$    &  $ $2&.3439$\dexp{}$  & $ $2&.3439$\dexp{}$	& $ $2&.3441$\dexp{}$  & $ $2&.3438$\dexp{}$  & $ $2&.3438$\dexp{}$\\
		\hline
		\hline
\end{longtable}
\newpage
\begin{longtable}{rlr@{}lr@{}lr@{}lr@{}lr@{}l}
\caption{\label{tab:g3_2p_12_12}
Third-order Zeeman effect for the ground state ($\stgr$, $M_J=1/2$) of boronlike ions in terms of the dimensionless coefficient $\gBBB{}$, see Eq.~(\ref{eq:g-3}). The leading-order term $\gBBB{0}$, the one-photon-exchange correction $\dgBBBinto$, and their sum $\gBBBint$ are given for different screening potentials: core-Hartree (CH), Kohn-Sham (KS), Dirac-Slater (DS), local Dirac-Fock (LDF), and Perdew-Zunger (PZ).
} 
\vspace{1cm}
    \\
		\hline
		\hline
		$Z$
		& 
		& \multicolumn{2}{c}{CH}
		& \multicolumn{2}{c}{KS}
		& \multicolumn{2}{c}{DS} 
		& \multicolumn{2}{c}{LDF}
		& \multicolumn{2}{c}{PZ} \\
		\hline
		& $\gBBB{0}$ &  2&.074$\dexp{14}$ & 1&.876$\dexp{14}$ & 1&.031$\dexp{14}$  & 2&.336$\dexp{14}$ & 1&.185$\dexp{14}$ \\
	6   & $\dgBBBinto$  &  0&.239$\dexp{14}$ & 0&.213$\dexp{14}$ & 1&.071$\dexp{14}$  & 0&.043$\dexp{14}$ & 1&.017$\dexp{14}$ \\
		& $\gBBBint$ &  2&.313$\dexp{14}$ & 2&.090$\dexp{14}$ & 2&.103$\dexp{14}$  & 2&.380$\dexp{14}$ & 2&.202$\dexp{14}$ \\
		\hline
		& $\gBBB{0}$ &  6&.231$\dexp{12}$ & 5&.784$\dexp{12}$ & 3&.982$\dexp{12}$  & 6&.746$\dexp{12}$ & 4&.399$\dexp{12}$ \\
	8   & $\dgBBBinto$  &  0&.964$\dexp{12}$ & 1&.170$\dexp{12}$ & 2&.715$\dexp{12}$  & 0&.538$\dexp{12}$ & 2&.494$\dexp{12}$ \\
		& $\gBBBint$ &  7&.195$\dexp{12}$ & 6&.955$\dexp{12}$ & 6&.698$\dexp{12}$  & 7&.284$\dexp{12}$ & 7&.894$\dexp{12}$ \\
		\hline
		& $\gBBB{0}$ &  5&.805$\dexp{11}$ & 5&.471$\dexp{11}$ & 4&.159$\dexp{11}$  & 6&.125$\dexp{11}$ & 4&.491$\dexp{11}$ \\
	10  & $\dgBBBinto$  &  0&.792$\dexp{11}$ & 1&.008$\dexp{11}$ & 2&.127$\dexp{11}$  & 0&.521$\dexp{11}$ & 1&.915$\dexp{11}$ \\
		& $\gBBBint$ &  6&.597$\dexp{11}$ & 6&.480$\dexp{11}$ & 6&.286$\dexp{11}$  & 6&.647$\dexp{11}$ & 6&.407$\dexp{11}$ \\
		\hline
		& $\gBBB{0}$ &  0&.947$\dexp{11}$ & 0&.902$\dexp{11}$ & 0&.725$\dexp{11}$  & 0&.984$\dexp{11}$ & 0&.772$\dexp{11}$ \\
	12  & $\dgBBBinto$  &  0&.111$\dexp{11}$ & 0&.145$\dexp{11}$ & 0&.297$\dexp{11}$  & 0&.079$\dexp{11}$ & 0&.264$\dexp{11}$ \\
		& $\gBBBint$ &  1&.058$\dexp{11}$ & 1&.047$\dexp{11}$ & 1&.023$\dexp{11}$  & 1&.064$\dexp{11}$ & 1&.037$\dexp{11}$ \\
		\hline
		& $\gBBB{0}$ &  2&.175$\dexp{10}$ & 2&.086$\dexp{10}$ & 1&.742$\dexp{10}$  & 2&.240$\dexp{10}$ & 1&.835$\dexp{10}$ \\
	14  & $\dgBBBinto$  &  0&.223$\dexp{10}$ & 0&.293$\dexp{10}$ & 0&.596$\dexp{10}$  & 0&.166$\dexp{10}$ & 0&.526$\dexp{10}$ \\
		& $\gBBBint$ &  2&.399$\dexp{10}$ & 2&.380$\dexp{10}$ & 2&.338$\dexp{10}$  & 2&.407$\dexp{10}$ & 2&.362$\dexp{10}$ \\
		\hline
		& $\gBBB{0}$ &  6&.298$\dexp{9}$  & 6&.073$\dexp{9}$  & 5&.206$\dexp{9}$   & 6&.447$\dexp{9}$  & 5&.445$\dexp{9}$ \\
	16  & $\dgBBBinto$  &  0&.572$\dexp{9}$  & 0&.756$\dexp{9}$  & 1&.530$\dexp{9}$   & 0&.440$\dexp{9}$  & 1&.342$\dexp{9}$ \\
		& $\gBBBint$ &  6&.870$\dexp{9}$  & 6&.830$\dexp{9}$  & 6&.736$\dexp{9}$   & 6&.888$\dexp{9}$  & 6&.787$\dexp{9}$ \\		
		\hline
		& $\gBBB{0}$ &  2&.156$\dexp{9}$  & 2&.088$\dexp{9}$  & 1&.825$\dexp{9}$   & 2&.198$\dexp{9}$  & 1&.898$\dexp{9}$ \\
	18  & $\dgBBBinto$  &  0&.175$\dexp{9}$  & 0&.232$\dexp{9}$  & 0&.470$\dexp{9}$   & 0&.138$\dexp{9}$  & 0&.410$\dexp{9}$ \\
		& $\gBBBint$ &  2&.331$\dexp{9}$  & 2&.321$\dexp{9}$  & 2&.295$\dexp{9}$   & 2&.336$\dexp{9}$  & 2&.309$\dexp{9}$ \\
		\hline
		& $\gBBB{0}$ &  8&.387$\dexp{8}$  & 8&.150$\dexp{8}$  & 7&.232$\dexp{8}$   & 8&.521$\dexp{8}$  & 7&.489$\dexp{8}$ \\
	20  & $\dgBBBinto$  &  0&.615$\dexp{8}$  & 0&.821$\dexp{8}$  & 1&.657$\dexp{8}$   & 0&.495$\dexp{8}$  & 1&.443$\dexp{8}$ \\
		& $\gBBBint$ &  9&.003$\dexp{8}$  & 8&.971$\dexp{8}$  & 8&.889$\dexp{8}$   & 9&.017$\dexp{8}$  & 8&.933$\dexp{8}$ \\
		\hline
		& $\gBBB{0}$ &  3&.605$\dexp{8}$  & 3&.513$\dexp{8}$  & 3&.155$\dexp{8}$   & 3&.654$\dexp{8}$  & 3&.255$\dexp{8}$ \\
	22  & $\dgBBBinto$  &  0&.241$\dexp{8}$  & 0&.322$\dexp{8}$  & 0&.651$\dexp{8}$   & 0&.197$\dexp{8}$  & 0&.566$\dexp{8}$ \\
		& $\gBBBint$ &  3&.846$\dexp{8}$  & 3&.836$\dexp{8}$  & 3&.806$\dexp{8}$   & 3&.851$\dexp{8}$  & 3&.822$\dexp{8}$ \\
		\hline
		& $\gBBB{0}$ &  1&.680$\dexp{8}$  & 1&.641$\dexp{8}$  & 1&.488$\dexp{8}$   & 1&.699$\dexp{8}$  & 1&.531$\dexp{8}$ \\
	24  & $\dgBBBinto$  &  0&.103$\dexp{8}$  & 0&.138$\dexp{8}$  & 0&.279$\dexp{8}$   & 0&.085$\dexp{8}$  & 0&.242$\dexp{8}$ \\
		& $\gBBBint$ &  1&.783$\dexp{8}$  & 1&.779$\dexp{8}$  & 1&.767$\dexp{8}$   & 1&.785$\dexp{8}$  & 1&.773$\dexp{8}$ \\
		\hline
		& $\gBBB{0}$ &  1&.398$\dexp{7}$  & 1&.375$\dexp{7}$  & 1&.280$\dexp{7}$   & 1&.408$\dexp{7}$  & 1&.306$\dexp{7}$ \\
	32  & $\dgBBBinto$  &  0&.065$\dexp{7}$  & 0&.087$\dexp{7}$  & 0&.176$\dexp{7}$   & 0&.056$\dexp{7}$  & 0&.152$\dexp{7}$ \\
		& $\gBBBint$ &  1&.464$\dexp{7}$  & 1&.462$\dexp{7}$  & 1&.456$\dexp{7}$   & 1&.464$\dexp{7}$  & 1&.459$\dexp{7}$ \\
		\hline
		& $\gBBB{0}$ &  1&.6599$\dexp{5}$ & 1&.6457$\dexp{5}$ & 1&.5778$\dexp{5}$  & 1&.6628$\dexp{5}$ & 1&.5945$\dexp{5}$\\
	54  & $\dgBBBinto$  &  0&.0470$\dexp{5}$ & 0&.0607$\dexp{5}$ & 0&.1264$\dexp{5}$  & 0&.0445$\dexp{5}$ & 0&.1105$\dexp{5}$\\
		& $\gBBBint$ &  1&.7069$\dexp{5}$ & 1&.7065$\dexp{5}$ & 1&.7042$\dexp{5}$  & 1&.7073$\dexp{5}$ & 1&.7051$\dexp{5}$\\
		\hline
		& $\gBBB{0}$ &  4&.5726$\dexp{3}$ & 4&.5533$\dexp{3}$ & 4&.4229$\dexp{3}$  & 4&.5650$\dexp{3}$ & 4&.4453$\dexp{3}$\\
	82  & $\dgBBBinto$  &  0&.0946$\dexp{3}$ & 0&.1141$\dexp{3}$ & 0&.2417$\dexp{3}$  & 0&.1028$\dexp{3}$ & 0&.2198$\dexp{3}$\\
		& $\gBBBint$ &  4&.6672$\dexp{3}$ & 4&.6674$\dexp{3}$ & 4&.6647$\dexp{3}$  & 4&.6679$\dexp{3}$ & 4&.6651$\dexp{3}$\\
		\hline
		& $\gBBB{0}$ &  4&.1035$\dexp{3}$ & 4&.0866$\dexp{3}$ & 3&.9706$\dexp{3}$  & 4&.0963$\dexp{3}$ & 3&.9901$\dexp{3}$\\
	83  & $\dgBBBinto$  &  0&.0844$\dexp{3}$ & 0&.1015$\dexp{3}$ & 0&.2151$\dexp{3}$  & 0&.0922$\dexp{3}$ & 0&.1959$\dexp{3}$\\
		& $\gBBBint$ &  4&.1879$\dexp{3}$ & 4&.1882$\dexp{3}$ & 4&.1858$\dexp{3}$  & 4&.1885$\dexp{3}$ & 4&.1861$\dexp{3}$\\
		\hline
		& $\gBBB{0}$ &  1&.6033$\dexp{3}$ & 1&.5978$\dexp{3}$ & 1&.5555$\dexp{3}$  & 1&.5987$\dexp{3}$ & 1&.5613$\dexp{3}$\\
	92  & $\dgBBBinto$  &  0&.0317$\dexp{3}$ & 0&.0373$\dexp{3}$ & 0&.0788$\dexp{3}$  & 0&.0365$\dexp{3}$ & 0&.0731$\dexp{3}$\\
		& $\gBBBint$ &  1&.6350$\dexp{3}$ & 1&.6352$\dexp{3}$ & 1&.6344$\dexp{3}$  & 1&.6352$\dexp{3}$ & 1&.6344$\dexp{3}$\\
		\hline
		\hline
\end{longtable}
\newpage
\begin{longtable}{rlr@{}lr@{}lr@{}lr@{}lr@{}l}
\caption{\label{tab:g3_2p_32_12}
Third-order Zeeman effect for the first excited state ($\stex$, $M_J=1/2$) of boronlike ions in terms of the dimensionless coefficient $\gBBB{}$, see Eq.~(\ref{eq:g-3}). The leading-order term $\gBBB{0}$, the one-photon-exchange correction $\dgBBBinto$, and their sum $\gBBBint$ are given for different screening potentials: core-Hartree (CH), Kohn-Sham (KS), Dirac-Slater (DS), local Dirac-Fock (LDF), and Perdew-Zunger (PZ).
} 
\vspace{1cm}
    \\
		\hline
		\hline
		$Z$
		& 
		& \multicolumn{2}{c}{CH}
		& \multicolumn{2}{c}{KS}
		& \multicolumn{2}{c}{DS} 
		& \multicolumn{2}{c}{LDF}
		& \multicolumn{2}{c}{PZ} \\
		\hline
	& $\gBBB{0}$ &  $-$2&.074$\dexp{14}$ & $-$1&.875$\dexp{14}$ & $-$1&.031$\dexp{14}$ & $-$2&.336$\dexp{14}$ & $-$1&.185$\dexp{14}$ \\
6   & $\dgBBBinto$  &  $-$0&.239$\dexp{14}$ & $-$0&.215$\dexp{14}$ & $-$1&.071$\dexp{14}$ & $-$0&.043$\dexp{14}$ & $-$1&.016$\dexp{14}$ \\
	& $\gBBBint$ &  $-$2&.313$\dexp{14}$ & $-$2&.091$\dexp{14}$ & $-$2&.102$\dexp{14}$ & $-$2&.380$\dexp{14}$ & $-$2&.202$\dexp{14}$ \\
		\hline
	& $\gBBB{0}$ &  $-$6&.231$\dexp{12}$ & $-$5&.780$\dexp{12}$ & $-$3&.980$\dexp{12}$ & $-$6&.746$\dexp{12}$ & $-$4&.399$\dexp{12}$ \\
8   & $\dgBBBinto$  &  $-$0&.964$\dexp{12}$ & $-$1&.175$\dexp{12}$ & $-$2&.716$\dexp{12}$ & $-$0&.538$\dexp{12}$ & $-$2&.494$\dexp{12}$ \\
	& $\gBBBint$ &  $-$7&.195$\dexp{12}$ & $-$6&.955$\dexp{12}$ & $-$6&.696$\dexp{12}$ & $-$7&.284$\dexp{12}$ & $-$6&.894$\dexp{12}$ \\
		\hline
	& $\gBBB{0}$ &  $-$5&.805$\dexp{11}$ & $-$5&.467$\dexp{11}$ & $-$4&.156$\dexp{11}$ & $-$6&.125$\dexp{11}$ & $-$4&.491$\dexp{11}$\\
10  & $\dgBBBinto$  &  $-$0&.792$\dexp{11}$ & $-$1&.012$\dexp{11}$ & $-$2&.128$\dexp{11}$ & $-$0&.521$\dexp{11}$ & $-$1&.915$\dexp{11}$\\
	& $\gBBBint$ &  $-$6&.597$\dexp{11}$ & $-$6&.480$\dexp{11}$ & $-$6&.285$\dexp{11}$ & $-$6&.647$\dexp{11}$ & $-$6&.407$\dexp{11}$\\
		\hline
	& $\gBBB{0}$ &  $-$0&.947$\dexp{11}$ & $-$0&.901$\dexp{11}$ & $-$0&.725$\dexp{11}$ & $-$0&.984$\dexp{11}$ & $-$0&.772$\dexp{11}$\\
12  & $\dgBBBinto$  &  $-$0&.111$\dexp{11}$ & $-$0&.145$\dexp{11}$ & $-$0&.297$\dexp{11}$ & $-$0&.079$\dexp{11}$ & $-$0&.264$\dexp{11}$\\
	& $\gBBBint$ &  $-$1&.059$\dexp{11}$ & $-$1&.047$\dexp{11}$ & $-$1&.023$\dexp{11}$ & $-$1&.064$\dexp{11}$ & $-$1&.037$\dexp{11}$\\
		\hline
	& $\gBBB{0}$ &  $-$2&.175$\dexp{10}$ & $-$2&.084$\dexp{10}$ & $-$1&.740$\dexp{10}$ & $-$2&.240$\dexp{10}$ & $-$1&.835$\dexp{10}$\\
14  & $\dgBBBinto$  &  $-$0&.223$\dexp{10}$ & $-$0&.295$\dexp{10}$ & $-$0&.597$\dexp{10}$ & $-$0&.166$\dexp{10}$ & $-$0&.526$\dexp{10}$\\
	& $\gBBBint$ &  $-$2&.399$\dexp{10}$ & $-$2&.380$\dexp{10}$ & $-$2&.338$\dexp{10}$ & $-$2&.407$\dexp{10}$ & $-$2&.362$\dexp{10}$\\
		\hline
	& $\gBBB{0}$ &  $-$6&.298$\dexp{9}$  & $-$6&.066$\dexp{9}$  & $-$5&.201$\dexp{9}$ &  $-$6&.447$\dexp{9}$  & $-$5&.445$\dexp{9}$ \\
16  & $\dgBBBinto$  &  $-$0&.572$\dexp{9}$  & $-$0&.763$\dexp{9}$  & $-$1&.534$\dexp{9}$ &  $-$0&.440$\dexp{9}$  & $-$1&.342$\dexp{9}$ \\
	& $\gBBBint$ &  $-$6&.870$\dexp{9}$  & $-$6&.830$\dexp{9}$  & $-$6&.735$\dexp{9}$ &  $-$6&.888$\dexp{9}$  & $-$6&.787$\dexp{9}$ \\		
		\hline
	& $\gBBB{0}$ &  $-$2&.156$\dexp{9}$  & $-$2&.085$\dexp{9}$  & $-$1&.823$\dexp{9}$ &  $-$2&.198$\dexp{9}$  & $-$1&.898$\dexp{9}$ \\
18  & $\dgBBBinto$  &  $-$0&.175$\dexp{9}$  & $-$0&.235$\dexp{9}$  & $-$0&.471$\dexp{9}$ &  $-$0&.138$\dexp{9}$  & $-$0&.411$\dexp{9}$ \\
	& $\gBBBint$ &  $-$2&.331$\dexp{9}$  & $-$2&.321$\dexp{9}$  & $-$2&.295$\dexp{9}$ &  $-$2&.336$\dexp{9}$  & $-$2&.309$\dexp{9}$ \\
		\hline
	& $\gBBB{0}$ &  $-$8&.387$\dexp{8}$  & $-$8&.139$\dexp{8}$  & $-$7&.223$\dexp{8}$ &  $-$8&.521$\dexp{8}$  & $-$7&.489$\dexp{8}$ \\
20  & $\dgBBBinto$  &  $-$0&.615$\dexp{8}$  & $-$0&.832$\dexp{8}$  & $-$1&.664$\dexp{8}$ &  $-$0&.495$\dexp{8}$  & $-$1&.443$\dexp{8}$ \\
	& $\gBBBint$ &  $-$9&.003$\dexp{8}$  & $-$8&.971$\dexp{8}$  & $-$8&.888$\dexp{8}$ &  $-$9&.017$\dexp{8}$  & $-$8&.933$\dexp{8}$ \\
		\hline
	& $\gBBB{0}$ &  $-$3&.605$\dexp{8}$  & $-$3&.508$\dexp{8}$  & $-$3&.151$\dexp{8}$ &  $-$3&.654$\dexp{8}$  & $-$3&.255$\dexp{8}$ \\
22  & $\dgBBBinto$  &  $-$0&.241$\dexp{8}$  & $-$0&.327$\dexp{8}$  & $-$0&.655$\dexp{8}$ &  $-$0&.197$\dexp{8}$  & $-$0&.566$\dexp{8}$ \\
	& $\gBBBint$ &  $-$3&.846$\dexp{8}$  & $-$3&.835$\dexp{8}$  & $-$3&.806$\dexp{8}$ &  $-$3&.851$\dexp{8}$  & $-$3&.822$\dexp{8}$ \\
		\hline
	& $\gBBB{0}$ &  $-$1&.680$\dexp{8}$  & $-$1&.638$\dexp{8}$  & $-$1&.486$\dexp{8}$ &  $-$1&.699$\dexp{8}$  & $-$1&.531$\dexp{8}$\\
24  & $\dgBBBinto$  &  $-$0&.103$\dexp{8}$  & $-$0&.140$\dexp{8}$  & $-$0&.281$\dexp{8}$ &  $-$0&.085$\dexp{8}$  & $-$0&.242$\dexp{8}$\\
	& $\gBBBint$ &  $-$1&.783$\dexp{8}$  & $-$1&.779$\dexp{8}$  & $-$1&.767$\dexp{8}$ &  $-$1&.785$\dexp{8}$  & $-$1&.773$\dexp{8}$\\
		\hline
	& $\gBBB{0}$ &  $-$1&.399$\dexp{7}$  & $-$1&.372$\dexp{7}$  & $-$1&.277$\dexp{7}$ &  $-$1&.408$\dexp{7}$  & $-$1&.306$\dexp{7}$\\
32  & $\dgBBBinto$  &  $-$0&.065$\dexp{7}$  & $-$0&.090$\dexp{7}$  & $-$0&.178$\dexp{7}$ &  $-$0&.056$\dexp{7}$  & $-$0&.152$\dexp{7}$\\
	& $\gBBBint$ &  $-$1&.464$\dexp{7}$  & $-$1&.462$\dexp{7}$  & $-$1&.456$\dexp{7}$ &  $-$1&.464$\dexp{7}$  & $-$1&.459$\dexp{7}$\\
		\hline
	& $\gBBB{0}$ &  $-$1&.6611$\dexp{5}$ & $-$1&.6405$\dexp{5}$ & $-$1&.5738$\dexp{5}$&  $-$1&.6640$\dexp{5}$ & $-$1&.5958$\dexp{5}$\\
54  & $\dgBBBinto$  &  $-$0&.0470$\dexp{5}$ & $-$0&.0671$\dexp{5}$ & $-$0&.1313$\dexp{5}$&  $-$0&.0444$\dexp{5}$ & $-$0&.1105$\dexp{5}$\\
	& $\gBBBint$ &  $-$1&.7081$\dexp{5}$ & $-$1&.7076$\dexp{5}$ & $-$1&.7051$\dexp{5}$&  $-$1&.7085$\dexp{5}$ & $-$1&.7063$\dexp{5}$\\
		\hline
	& $\gBBB{0}$ &  $-$4&.6062$\dexp{3}$ & $-$4&.5554$\dexp{3}$ & $-$4&.4298$\dexp{3}$&  $-$4&.5986$\dexp{3}$ & $-$4&.4785$\dexp{3}$\\
82  & $\dgBBBinto$  &  $-$0&.0941$\dexp{3}$ & $-$0&.1449$\dexp{3}$ & $-$0&.2668$\dexp{3}$&  $-$0&.1024$\dexp{3}$ & $-$0&.2198$\dexp{3}$\\
	& $\gBBBint$ &  $-$4&.7004$\dexp{3}$ & $-$4&.7003$\dexp{3}$ & $-$4&.6967$\dexp{3}$&  $-$4&.7011$\dexp{3}$ & $-$4&.6983$\dexp{3}$\\
		\hline
	& $\gBBB{0}$ &  $-$4&.1355$\dexp{3}$ & $-$4&.0897$\dexp{3}$ & $-$3&.9781$\dexp{3}$&  $-$4&.1282$\dexp{3}$ & $-$4&.0217$\dexp{3}$ \\
83  & $\dgBBBinto$  &  $-$0&.0839$\dexp{3}$ & $-$0&.1297$\dexp{3}$ & $-$0&.2381$\dexp{3}$&  $-$0&.0918$\dexp{3}$ & $-$0&.1959$\dexp{3}$ \\
	& $\gBBBint$ &  $-$4&.2195$\dexp{3}$ & $-$4&.2194$\dexp{3}$ & $-$4&.2162$\dexp{3}$&  $-$4&.2201$\dexp{3}$ & $-$4&.2177$\dexp{3}$ \\
		\hline
	& $\gBBB{0}$ &  $-$1&.6228$\dexp{3}$ & $-$1&.6036$\dexp{3}$ & $-$1&.5632$\dexp{3}$&  $-$1&.6181$\dexp{3}$ & $-$1&.5805$\dexp{3}$ \\
92  & $\dgBBBinto$  &  $-$0&.0315$\dexp{3}$ & $-$0&.0508$\dexp{3}$ & $-$0&.0901$\dexp{3}$&  $-$0&.0365$\dexp{3}$ & $-$0&.0733$\dexp{3}$ \\
	& $\gBBBint$ &  $-$1&.6544$\dexp{3}$ & $-$1&.6545$\dexp{3}$ & $-$1&.6533$\dexp{3}$&  $-$1&.6546$\dexp{3}$ & $-$1&.6538$\dexp{3}$ \\
		\hline
		\hline
\end{longtable}
\newpage
\begin{longtable}{rlr@{}lr@{}lr@{}lr@{}lr@{}l}
\caption{\label{tab:g3_2p_32_32}
Third-order Zeeman effect for the first excited state ($\stex$, $M_J=3/2$) of boronlike ions in terms of the dimensionless coefficient $\gBBB{}$, see Eq.~(\ref{eq:g-3}). The leading-order term $\gBBB{0}$, the one-photon-exchange correction $\dgBBBinto$, and their sum $\gBBBint$ are given for different screening potentials: core-Hartree (CH), Kohn-Sham (KS), Dirac-Slater (DS), local Dirac-Fock (LDF), and Perdew-Zunger (PZ).
}
\vspace{1cm}
    \\
		\hline
		\hline
		$Z$
		& 
		& \multicolumn{2}{c}{CH}
		& \multicolumn{2}{c}{KS}
		& \multicolumn{2}{c}{DS} 
		& \multicolumn{2}{c}{LDF}
		& \multicolumn{2}{c}{PZ} \\
		\hline
	& $\gBBB{0}$ & $-$1&.742$\dexp{5}$   & $-$2&.275$\dexp{5}$& $-$1&.719$\dexp{5}$ & $-$1&.704$\dexp{5}$ & $-$1&.436$\dexp{5}$ \\
6   & $\dgBBBinto$  &    0&.564$\dexp{5}$   &    1&.540$\dexp{5}$&    0&.373$\dexp{5}$ &    0&.448$\dexp{5}$ &    0&.036$\dexp{5}$ \\
	& $\gBBBint$ & $-$1&.178$\dexp{5}$   & $-$0&.734$\dexp{5}$& $-$1&.346$\dexp{5}$ & $-$1&.256$\dexp{5}$ & $-$1&.399$\dexp{5}$ \\
		\hline
	& $\gBBB{0}$ & $-$5&.375$\dexp{4}$   & $-$6&.208$\dexp{4}$& $-$5&.290$\dexp{4}$  & $-$5&.414$\dexp{4}$ & $-$4&.900$\dexp{4}$ \\
8   & $\dgBBBinto$  &    0&.871$\dexp{4}$   &    2&.064$\dexp{4}$&    0&.615$\dexp{4}$  &    0&.857$\dexp{4}$ &    0&.175$\dexp{4}$ \\
	& $\gBBBint$ & $-$4&.503$\dexp{4}$   & $-$4&.143$\dexp{4}$& $-$4&.675$\dexp{4}$  & $-$4&.556$\dexp{4}$ & $-$4&.724$\dexp{4}$ \\
		\hline
	& $\gBBB{0}$ & $-$2&.643$\dexp{4}$   & $-$2&.916$\dexp{4}$& $-$2&.605$\dexp{4}$  & $-$2&.669$\dexp{4}$ & $-$2&.486$\dexp{4}$ \\
10  & $\dgBBBinto$  &    0&.292$\dexp{4}$   &    0&.645$\dexp{4}$&    0&.210$\dexp{4}$  &    0&.306$\dexp{4}$ &    0&.080$\dexp{4}$ \\
	& $\gBBBint$ & $-$2&.351$\dexp{4}$   & $-$2&.271$\dexp{4}$& $-$2&.394$\dexp{4}$  & $-$2&.362$\dexp{4}$ & $-$2&.406$\dexp{4}$ \\
		\hline
	& $\gBBB{0}$ & $-$1&.577$\dexp{4}$   & $-$1&.699$\dexp{4}$& $-$1&.557$\dexp{4}$  & $-$1&.591$\dexp{4}$ & $-$1&.506$\dexp{4}$ \\
12  & $\dgBBBinto$  &    0&.133$\dexp{4}$   &    0&.281$\dexp{4}$&    0&.097$\dexp{4}$  &    0&.143$\dexp{4}$ &    0&.042$\dexp{4}$ \\
	& $\gBBBint$ & $-$1&.444$\dexp{4}$   & $-$1&.417$\dexp{4}$& $-$1&.460$\dexp{4}$  & $-$1&.448$\dexp{4}$ & $-$1&.464$\dexp{4}$ \\
		\hline
	& $\gBBB{0}$ & $-$1&.048$\dexp{4}$   & $-$1&.113$\dexp{4}$& $-$1&.037$\dexp{4}$  & $-$1&.057$\dexp{4}$ & $-$1&.011$\dexp{4}$ \\
14  & $\dgBBBinto$  &    0&.071$\dexp{4}$   &    0&.147$\dexp{4}$&    0&.052$\dexp{4}$  &    0&.078$\dexp{4}$ &    0&.025$\dexp{4}$ \\
	& $\gBBBint$ & $-$0&.977$\dexp{4}$   & $-$0&.965$\dexp{4}$& $-$0&.984$\dexp{4}$  & $-$0&.978$\dexp{4}$ & $-$0&.986$\dexp{4}$ \\
		\hline
	& $\gBBB{0}$ & $-$7&.482$\dexp{3}$   & $-$7&.868$\dexp{3}$& $-$7&.406$\dexp{3}$  & $-$7&.538$\dexp{3}$ & $-$7&.256$\dexp{3}$ \\
16  & $\dgBBBinto$  &    0&.430$\dexp{3}$   &    0&.872$\dexp{3}$&    0&.316$\dexp{3}$  &    0&.477$\dexp{3}$ &    0&.158$\dexp{3}$ \\
	& $\gBBBint$ & $-$7&.052$\dexp{3}$   & $-$6&.996$\dexp{3}$& $-$7&.090$\dexp{3}$  & $-$7&.061$\dexp{3}$ & $-$7&.098$\dexp{3}$ \\
		\hline
	& $\gBBB{0}$ & $-$5&.606$\dexp{3}$   & $-$5&.853$\dexp{3}$& $-$5&.555$\dexp{3}$  & $-$5&.644$\dexp{3}$ & $-$5&.460$\dexp{3}$ \\
18  & $\dgBBBinto$  &    0&.278$\dexp{3}$   &    0&.556$\dexp{3}$&    0&.205$\dexp{3}$  &    0&.310$\dexp{3}$ &    0&.106$\dexp{3}$ \\
	& $\gBBBint$ & $-$5&.328$\dexp{3}$   & $-$5&.297$\dexp{3}$& $-$5&.349$\dexp{3}$  & $-$5&.333$\dexp{3}$ & $-$5&.354$\dexp{3}$ \\
		\hline
	& $\gBBB{0}$ &  $-$4&.356$\dexp{3}$  & $-$4&.524$\dexp{3}$& $-$4&.320$\dexp{3}$  & $-$4&.383$\dexp{3}$ & $-$4&.257$\dexp{3}$ \\
20  & $\dgBBBinto$  &     0&.190$\dexp{3}$  &    0&.375$\dexp{3}$&    0&.140$\dexp{3}$  &    0&.213$\dexp{3}$ &    0&.074$\dexp{3}$ \\
	& $\gBBBint$ &  $-$4&.166$\dexp{3}$  & $-$4&.149$\dexp{3}$& $-$4&.180$\dexp{3}$  & $-$4&.170$\dexp{3}$ & $-$4&.183$\dexp{3}$ \\
		\hline
	& $\gBBB{0}$ &  $-$3&.482$\dexp{3}$  & $-$3&.601$\dexp{3}$& $-$3&.455$\dexp{3}$  & $-$3&.501$\dexp{3}$ & $-$3&.411$\dexp{3}$ \\
22  & $\dgBBBinto$  &     0&.135$\dexp{3}$  &    0&.265$\dexp{3}$&    0&.099$\dexp{3}$  &    0&.152$\dexp{3}$ &    0&.053$\dexp{3}$ \\
	& $\gBBBint$ &  $-$3&.346$\dexp{3}$  & $-$3&.335$\dexp{3}$& $-$3&.355$\dexp{3}$  & $-$3&.349$\dexp{3}$ & $-$3&.357$\dexp{3}$ \\
		\hline
	& $\gBBB{0}$ &  $-$2&.846$\dexp{3}$  & $-$2&.933$\dexp{3}$& $-$2&.825$\dexp{3}$  & $-$2&.860$\dexp{3}$ & $-$2&.793$\dexp{3}$ \\
24  & $\dgBBBinto$  &     0&.099$\dexp{3}$  &    0&.194$\dexp{3}$&    0&.073$\dexp{3}$  &    0&.112$\dexp{3}$ &    0&.040$\dexp{3}$ \\
    & $\gBBBint$ &  $-$2&.746$\dexp{3}$  & $-$2&.739$\dexp{3}$& $-$2&.752$\dexp{3}$  & $-$2&.747$\dexp{3}$ & $-$2&.753$\dexp{3}$ \\
		\hline
	& $\gBBB{0}$ &  $-$1&.482$\dexp{3}$  & $-$1&.515$\dexp{3}$& $-$1&.474$\dexp{3}$  & $-$1&.488$\dexp{3}$ & $-$1&.462$\dexp{3}$ \\
32  & $\dgBBBinto$  &     0&.036$\dexp{3}$  &    0&.071$\dexp{3}$&    0&.026$\dexp{3}$  &    0&.041$\dexp{3}$ &    0&.014$\dexp{3}$ \\
	& $\gBBBint$ &  $-$1&.445$\dexp{3}$  & $-$1&.443$\dexp{3}$& $-$1&.447$\dexp{3}$  & $-$1&.446$\dexp{3}$ & $-$1&.447$\dexp{3}$ \\
		\hline
	& $\gBBB{0}$ &  $-$4&.6571$\dexp{2}$ & $-$4&.7139$\dexp{2}$ & $-$4&.6400$\dexp{2}$ & $-$4&.6652$\dexp{2}$ & $-$4&.6189$\dexp{2}$ \\
54  & $\dgBBBinto$  &     0&.0609$\dexp{2}$ &    0&.1193$\dexp{2}$ &    0&.0418$\dexp{2}$ &    0&.0685$\dexp{2}$ &    0&.0206$\dexp{2}$ \\
	& $\gBBBint$ &  $-$4&.5961$\dexp{2}$ & $-$4&.5946$\dexp{2}$ & $-$4&.5982$\dexp{2}$ & $-$4&.5967$\dexp{2}$ & $-$4&.5983$\dexp{2}$ \\
		\hline
	& $\gBBB{0}$ &  $-$1&.8207$\dexp{2}$ & $-$1&.8346$\dexp{2}$ & $-$1&.8154$\dexp{2}$ & $-$1&.8215$\dexp{2}$ & $-$1&.8096$\dexp{2}$ \\
82  & $\dgBBBinto$  &     0&.0133$\dexp{2}$ &    0&.0274$\dexp{2}$ &    0&.0076$\dexp{2}$ &    0&.0140$\dexp{2}$ &    0&.0018$\dexp{2}$ \\
	& $\gBBBint$ &  $-$1&.8073$\dexp{2}$ & $-$1&.8072$\dexp{2}$ & $-$1&.8078$\dexp{2}$ & $-$1&.8075$\dexp{2}$ & $-$1&.8077$\dexp{2}$ \\
		\hline
	& $\gBBB{0}$ &  $-$1&.7703$\dexp{2}$ & $-$1&.7837$\dexp{2}$ & $-$1&.7653$\dexp{2}$ & $-$1&.7711$\dexp{2}$ & $-$1&.7596$\dexp{2}$\\
83  & $\dgBBBinto$  &     0&.0127$\dexp{2}$ &    0&.0262$\dexp{2}$ &    0&.0072$\dexp{2}$ &    0&.0133$\dexp{2}$ &    0&.0016$\dexp{2}$\\
	& $\gBBBint$ &  $-$1&.7576$\dexp{2}$ & $-$1&.7575$\dexp{2}$ & $-$1&.7580$\dexp{2}$ & $-$1&.7577$\dexp{2}$ & $-$1&.7580$\dexp{2}$\\
		\hline
	& $\gBBB{0}$ &  $-$1&.3910$\dexp{2}$ & $-$1&.4003$\dexp{2}$ & $-$1&.3871$\dexp{2}$ & $-$1&.3911$\dexp{2}$ & $-$1&.3829$\dexp{2}$ \\
92  & $\dgBBBinto$  &     0&.0085$\dexp{2}$ &    0&.0179$\dexp{2}$ &    0&.0043$\dexp{2}$ &    0&.0085$\dexp{2}$ &    0&.0002$\dexp{2}$ \\
	& $\gBBBint$ &  $-$1&.3824$\dexp{2}$ & $-$1&.3824$\dexp{2}$ & $-$1&.3827$\dexp{2}$ & $-$1&.3825$\dexp{2}$ & $-$1&.3827$\dexp{2}$ \\
		\hline
		\hline
\end{longtable}
\newpage
\begin{table}
\caption{\label{tab:delta_g3}
The third-order Zeeman effect in boronlike ions represented as the $g$-factor correction $\delta g$ at the field of 1 Tesla, see Eq.~(\ref{eq:g-g-3}). The value of $\gBBBint$ for the local Dirac-Fock potential is used. The $g$-factor correction $\delta g$ at the field of $\beta$ Tesla can be found with multiplication by $\beta^2$.
}
\vspace{1cm}
	\begin{tabular}{rr@{}lr@{}lr@{}l}
		\hline
		\hline
		$Z$
		& \multicolumn{2}{c}{$\stgr$}
		& \multicolumn{4}{c}{$\stex$}
	 \\
		& \multicolumn{2}{c}{$M_J=\pm 1/2$}
		& \multicolumn{2}{c}{$M_J=\pm 1/2$}
		& \multicolumn{2}{c}{$M_J=\pm 3/2$}
	 \\
		\hline
	6   &  $ $6&.108$\dexp{-6}$   & $-$6&.108$\dexp{-6}$	& $-$1&.074$\dexp{-15}$  \\
	8   &  $ $1&.869$\dexp{-7}$   & $-$1&.869$\dexp{-7}$	& $-$3&.897$\dexp{-16}$  \\
   10   &  $ $1&.706$\dexp{-8}$   & $-$1&.706$\dexp{-8}$	& $-$2&.021$\dexp{-16}$  \\
   12   &  $ $2&.731$\dexp{-9}$   & $-$2&.731$\dexp{-9}$	& $-$1&.239$\dexp{-16}$  \\
   14   &  $ $6&.177$\dexp{-10}$  & $-$6&.177$\dexp{-10}$	& $-$8&.366$\dexp{-17}$  \\
   16   &  $ $1&.768$\dexp{-10}$  & $-$1&.768$\dexp{-10}$	& $-$6&.040$\dexp{-17}$  \\
   18   &  $ $5&.995$\dexp{-11}$  & $-$5&.995$\dexp{-11}$	& $-$4&.562$\dexp{-17}$  \\
   20   &  $ $2&.314$\dexp{-11}$  & $-$2&.314$\dexp{-11}$	& $-$3&.567$\dexp{-17}$  \\
   22   &  $ $9&.883$\dexp{-12}$  & $-$9&.883$\dexp{-12}$	& $-$2&.865$\dexp{-17}$  \\
   24  &   $ $4&.581$\dexp{-12}$  & $-$4&.581$\dexp{-12}$	& $-$2&.350$\dexp{-17}$  \\
   32  &   $ $3&.757$\dexp{-13}$  & $-$3&.757$\dexp{-13}$	& $-$1&.237$\dexp{-17}$  \\
   54   &  $ $4&.381$\dexp{-15}$  & $-$4&.385$\dexp{-15}$	& $-$3&.932$\dexp{-18}$  \\
   82   &  $ $1&.198$\dexp{-16}$  & $-$1&.206$\dexp{-16}$	& $-$1&.546$\dexp{-18}$  \\
   83   &  $ $1&.075$\dexp{-16}$  & $-$1&.083$\dexp{-16}$	& $-$1&.504$\dexp{-18}$  \\
   92   &  $ $4&.194$\dexp{-17}$  & $-$4&.246$\dexp{-17}$	& $-$1&.183$\dexp{-18}$  \\
		\hline
		\hline
	\end{tabular}
\end{table}
%
%
\end{document}